# A Systems Engineering Approach to Modeling and Analysis of Chronic Obstructive Pulmonary Disease (COPD)

**Varghese Kurian, Navid Ghadipasha, Michelle Gee, Anais Chalant, Teresa Hamill, Alphonse Okossi, Lucy Chen, Bin Yu, Babatunde A. Ogunnaike, Antony N. Beris**

***Abstract*— Chronic Obstructive Pulmonary Disease (COPD) is a progressive lung disease characterized by airflow limitation. This study develops a systems engineering framework for representing important mechanistic details of COPD in a model of the cardio-respiratory system. In this model, we present the cardio-respiratory system as an integrated biological control system responsible for regulating breathing. The four engineering control system components are considered: sensor, controller, actuator, and the process itself. Knowledge of human anatomy and physiology is used to develop appropriate mechanistic mathematical models for each component. Following a systematic analysis of the computational model, we identify three physiological parameters associated with reproducing clinical manifestations of COPD - changes in the forced expiratory volume (FEV), lung volumes, and pulmonary hypertension. We quantify the changes in these parameters (airway resistance, lung elastance, and pulmonary resistance) as the ones that result in a systemic response that is diagnostic of COPD. A multivariate analysis reveals that the changes in airway resistance have a broad impact on the human cardio-respiratory system, and that the pulmonary circuit is stressed beyond normal under hypoxic environments in most COPD patients.***

***Index Terms*—systems biology, COPD, hypoxia, human cardio-respiratory system, mathematical modeling, spirometry**

## I. Introduction

*‘Chronic Obstructive Pulmonary Disease (COPD) is a common, preventable, and treatable disease that is characterized by persistent respiratory symptoms and airflow limitation …*’[1] The airflow limitation that is due to either or a combination of bronchial and alveolar abnormalities is usually caused by significant exposure to noxious particles and gases such as cigarette smoke[2]. Chronic cough, excessive sputum production, and dyspnea are the most prevalent symptoms in COPD patients[3]. In general, patients experience a poorer quality of life and an increased risk of death[4]. Recent estimates of COPD indicate a global prevalence of about 40 crores (0.4 billion, 11.7% of population with age ≥ 30), and annual deaths of over 30 lakhs (3 million) [5,6]. Despite being a major public health challenge, there exist several impediments in the recognition, assessment, and management of the condition[7,8].

There are three common manifestations of COPD - *small airways disease, emphysema, and pulmonary hypertension (PH)* [1,9]. Patients with small airways disease are faced with difficulty in exhaling, which progressively traps gas in the lungs causing dynamic hyperinflation. This increase in lung volume is often associated with increased dyspnoea and exercise limitation[1]. Emphysema is an abnormal permanent enlargement of air sacs distal to the terminal bronchioles in the lungs. This enlargement can destroy the airspace walls, without obvious pulmonary fibrosis (i.e., there is no fibrosis visible to the naked eye)[10]. Additionally, in COPD patients, significant abnormalities are observed in the microvascular blood flow [9], which cause an increased pulmonary blood pressure resulting in PH. The stress associated with PH can lead to right ventricular hypertrophy (increase in the muscle mass of right ventricle) and eventually, cardiac failure. The relative contributions of these disease manifestations vary from person to person and may also evolve at different rates over time. This heterogeneity in disease trajectories along with the variations in response to therapy has promoted the emergence of *personalized medicine*[11] as an effective tool for the management of COPD[12]. However, at present, there are several impediments to this including the limitations in our understanding of the disease pathophysiology, and the lack of biomarkers[12].

Although there is no single definitive test for the diagnosis of COPD, spirometry is a common method which is used to measure airflow obstruction in COPD patients. Despite its high sensitivity in diagnosing COPD, spirometry cannot be reliably used as the only diagnostic test because of its weak specificity[13]. For this reason, other parameters such as symptoms and risk factors are also considered in conjunction with spirometric data. Efforts are also being made to evaluate the usefulness of radiographic measurements in the diagnosis

[1] This work is dedicated to the memory of Prof. Babatunde A. Ogunnaike who passed away on February 20, 2022. This work was supported in part by Air Liquide, by the Delaware Biotechnology Institute (Grant #12A00448), and through the use of DARWIN computing system at the University of Delaware (NSF Grant 1919839, R. Eigenmann, B. E. Bagozzi, A. Jayaraman, W. Totten, and C. H. Wu).

V Kurian and N Ghadipasha contributed equally to this work. V Kurian, M Gee, B Ogunnaike, and A Beris are with the Department of Chemical and Biomolecular Engineering at the University of Delaware, DE, USA. A Chalant, A Okossi, and L Chen, are with American Air Liquide Inc., Innovation Campus Delaware, Newark, DE, USA. N Ghadipasha was with the University of Delaware and T Hamill and B Yu were with Air Liquide when they contributed to the work. Correspondence concerning this article should be addressed to Antony N. Beris (beris@udel.edu).

of COPD, and to identify any biomarkers that are predictive/diagnostic of COPD[14].

*Mathematical models* are useful in analyzing the information contained in the physiological measurements and making appropriate inferences on the disease state of the patients. In this way, the models could assist in the personalization of COPD management strategies and help realize the clinical and economic benefits of *remote patient monitoring* – the collection and secure transmission of health data from individuals in one location to healthcare providers in a different location for assessment and recommendations [15]. Though there have been a few recent efforts towards developing models of the cardio-respiratory system[16-20], these have not been adapted to capture the response of COPD patients. For example, the model by Gutta et al.[20,21] uses reduced equations to represent certain aspects of the respiratory mechanics, the details of which are critical for representing the adaptations in COPD patients (but not for studying sleep apnea, the original application of the model). A more detailed model by Albanese et al.[16] assumes that the volume of air inhaled and exhaled in a respiratory cycle are the same, making it difficult to model the dynamics of air entrapment, a common symptom in COPD. On the other hand, the available models of COPD are either overly simplified, often ignoring important physical phenomena[22] or excessively detailed, making it almost impossible to use them at a systemic scale[23]. To the best of our knowledge, there exists no mathematical model of the human cardio-respiratory system that can provide insights into and make predictions of the systemic response of COPD patients to external stimuli.

We seek to bridge some of the gaps mentioned above by developing a physiology-based model of appropriate components of the cardio-respiratory system, using principles of systems and control engineering. Specifically, we propose modeling the occurrence of COPD from a control engineering perspective, whereby the cardio-respiratory system is represented as control system components whose physiological functions will be represented by appropriate mathematical equations (Section II). The performance of the proposed cardio-respiratory model is tested through some simulation case studies to ensure the accuracy and credibility of the model (Section III). In the next section (Section IV), a list of COPD-related parameters that are associated with the manifestations of COPD are identified and quantified. In Section V, a systematic analysis of the effect of changes in the parameters on cardio-respiratory variables is investigated. Finally, in Section VI, we present our conclusions.

## II. A Control Engineering Model of the Cardio-Respiratory System

Understanding the dynamics of the respiratory and cardiovascular systems, as well as their interactions, is essential to understand the underlying mechanisms of COPD. There is an increasing body of evidence that the inflammation associated with COPD is not limited to the lungs but can also affect non-pulmonary organs; in particular, the heart [24,25]. Therefore, the model should provide adequate physiological insights into both cardiovascular and respiratory systems.

Figure 1 shows a schematic of respiration and blood circulation in the body. As the blood moves from the lungs to the heart, then to the systemic networks, the oxygen concentration decreases, and the $CO_2$ concentration increases. Those are rectified in the lungs through the breathing process.

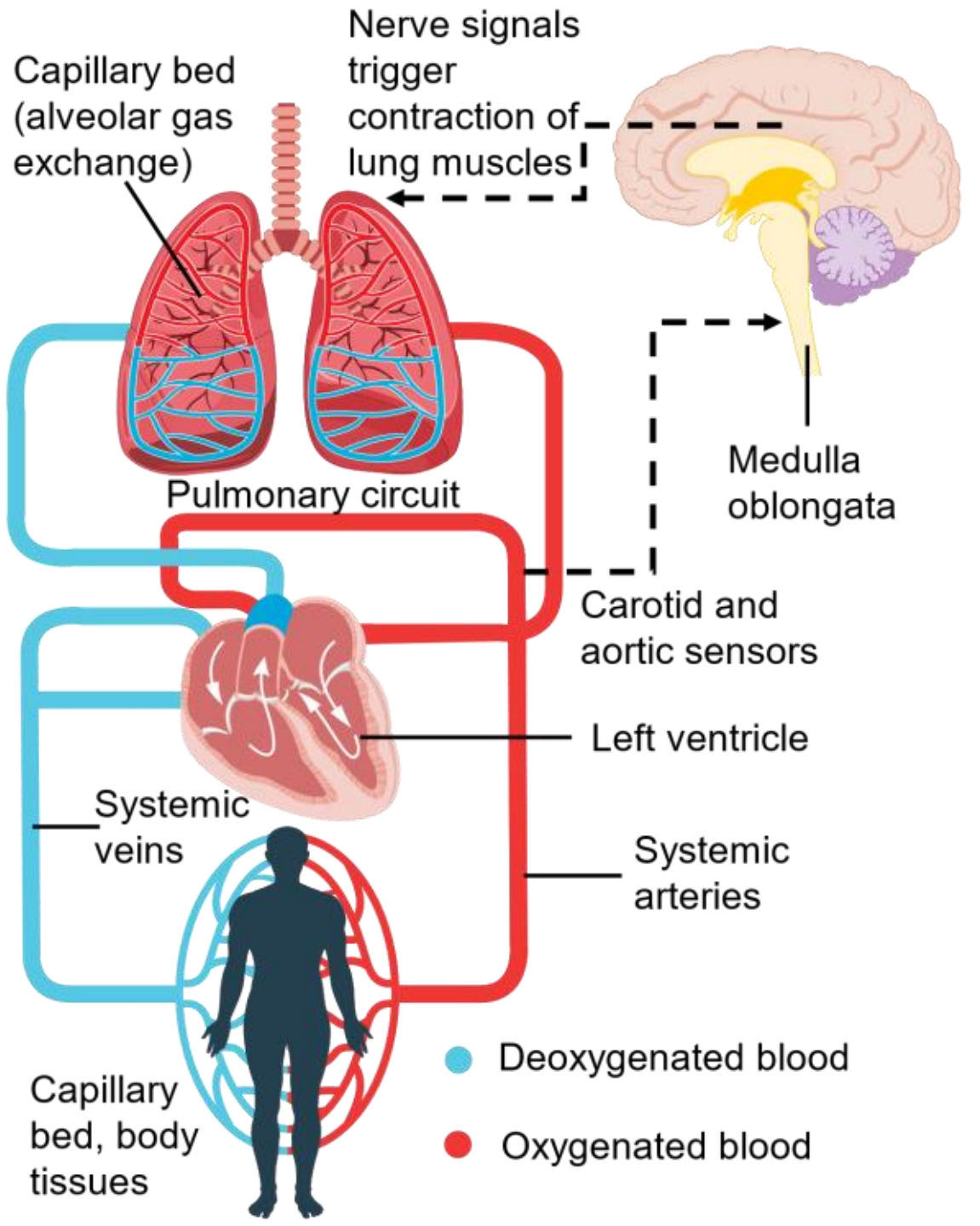

*Figure 1 Schematic of physiological components and mechanisms involved in breathing regulation and blood circulation. (Image of brain: Cancer Research UK; obtained via Wikimedia Commons. Image of cardio-respiratory system: istockphoto.com/colematt)*

Conceptually, the regulation of breathing is affected by an inherent biological control system consisting of complex interactions between the cardio-respiratory centers in the brain[26-28]. The harmonious interactions of these individual components generate the breathing rhythm and regulate the levels of oxygen and carbon dioxide in the body[29]. When the oxygenated blood flows through the systemic arteries, blood gas level (concentration of oxygen and carbon dioxide) is measured by the respiratory sensors on the arteries. Respiratory sensors are chemoreceptors, which can detect changes in the chemical concentrations of blood gas level. The chemoreceptors send the measured values of blood gas level as an electrical signal to the respiratory control center, which is located in the medulla oblongata in the brainstem. Based on the measured values, the controller generates another electrical signal which determines the contraction of the lung muscles. The lung muscles apply the pleural pressure, adjusting the breathing frequency and magnitude.

Figure 2 shows a control engineering representation of the above-mentioned components in the form of modules/blocks that perform the physiological functions of the sensor, controller, actuator, and process itself. This representation is particularly useful for the clarity with which it shows how each subsystem performs its function and how the various subsystems are connected such that the response ("output") of one subsystem provides the stimulus ("input") to another in the feedback loop. Specifically, the representation shows: 1) the *process:* how the "controlled variables" (or CVs, the physiological variables desired to be controlled), blood partial pressure of oxygen and carbon dioxide in the blood, are influenced by changes in the "manipulated variable" (MV), pleural pressure; 2) the *actuator*: how the MV—pleural pressure—in turn is determined by the "control action"—the electrical signal from the neural controller determines the contraction of the lung muscles; 3) the *controller*: how the control action signal to the lungs is determined in the control center in the brain, in response to the difference between the actual measured values of each of the CVs and their corresponding desired values; 4) the *sensor*: how the measurements of the CVs are determined in the peripheral and central chemoreceptors, in response to changes in these CVs, and thereby closing the loop.

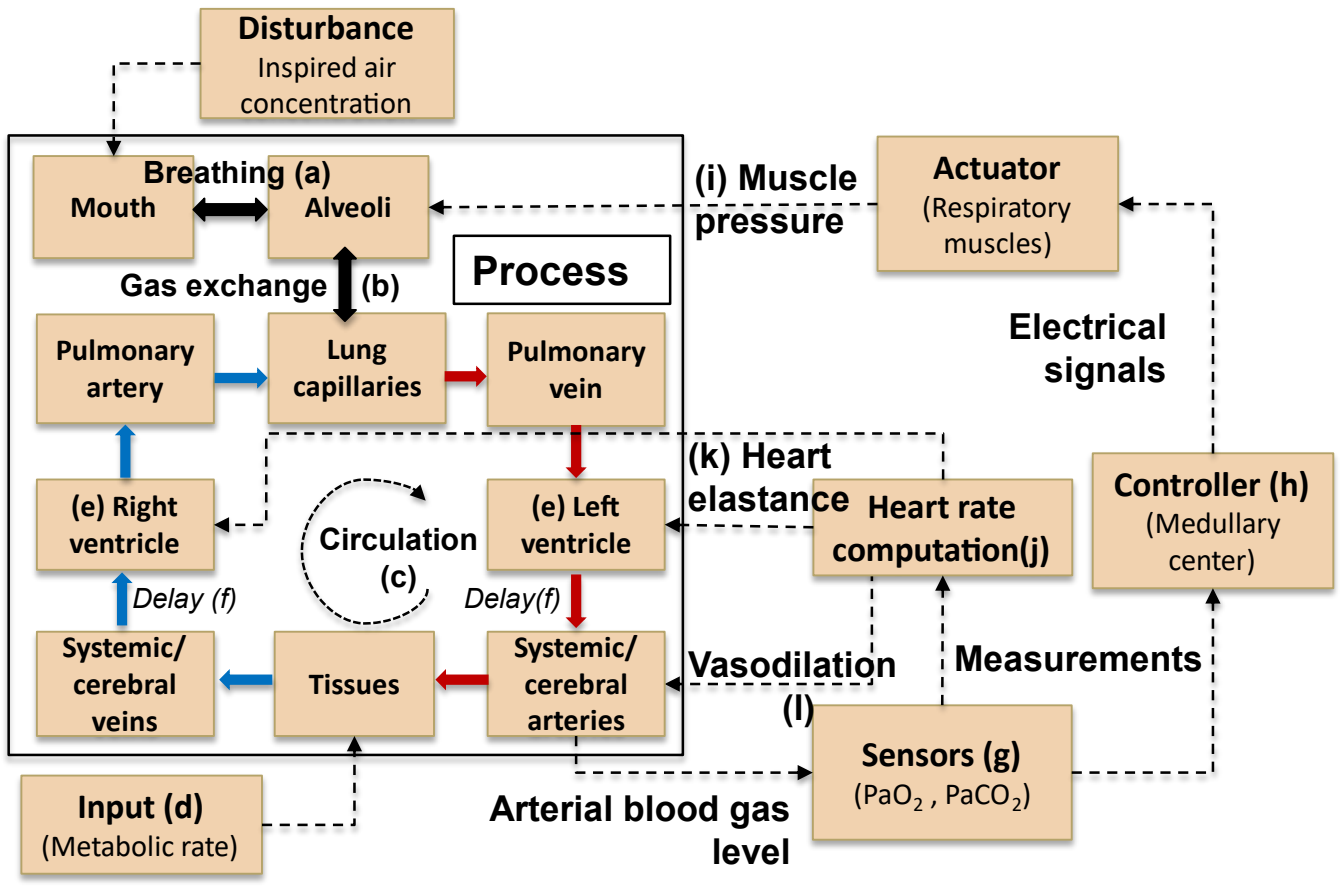

*Figure 2 Control engineering block diagram representation of the cardio-respiratory system. Dashed lines describe the flow of information while the solid black, blue, and red lines are the flow of gas, deoxygenated, and oxygenated blood, respectively.*

The mathematical expressions representing the mechanisms involved in the physiological function of each component module are derived from fundamental principles of material and momentum balance. The resulting differential-algebraic equations (summarized in Table 1, full set of equations available in Supplementary Information at the end of the document) show the output (response) of each block as a function of the corresponding input (stimulus) explicitly. In addition to the noted process variables, the model also consists of fixed parameters associated with the physiological characteristics of each component. The parameter values used in this work are either taken directly from literature or estimated using the input-output data of the corresponding physiological process. In reality, the specific values taken by these parameters depend on the patient in question and contribute to the model prediction of the response to any specified stimulus. Hence, the model response reported in this work is representative of the population-average rather than a specific individual (see [30] for a technique to generate a patient ensemble from the existing parameters). It may be noted that by selecting appropriate values for the physiological parameters, the model can be customized for individual patients.

The integrated cardio-respiratory model is a system of differential-algebraic equations with a few algebraic constraints. The model was developed using Simulink/MATLAB R2020b and solved using the MATLAB function ode23 that uses the Bockagi-Shampine method with a relative tolerance of $10^{-6}$ [31].

*Table 1 The functional forms used for representing key physiological processes and the main references. The row numbers correspond to the symbols used in Figure 2. Due to common original source(s), other models could have similarity in the functional forms we use.*

| | Physiological process | Model equation | Eq No. | Main source |
|---|---|---|---|---|
| a | Ventilation | Pressure driven flow | S.1 – 4 | [32,33] |
| b | Gas exchange in lungs | Diffusive transport | S.5 – 15 | [32-34] |
| c | Blood circulation | R-C circuit | S.16 –19 | [34] |
| d | Metabolism | $O_2$ and $CO_2$ mass balance | S.20 –21 | [34] |
| e | Ventricle mixing | CSTR | S.22 | [35] |
| f | Circulatory delay | Flow delay | S.23 | [36,37] |
| g | Sensors | Noiseless linear relation | S.24 –26 | [33,38] |
| h | Respiratory control | Ben-Tal method | S.27 –36 | [33,39] |
| i | Displacement of lung muscles | Linear ODE | S.37 –38 | [32,33] |
| j | Heart rate | Polynomial | S.39 –40 | [40] |
| k | Elastance of heart muscles | Trigonometric functions | S.41 | [34] |
| l | Vasodilation | Sigmoid of heart rate | S.42 | [41,42] |

## III. Simulation Results – Healthy Individuals

### *A. Stationary simulation results under standard conditions*

The first step towards verification of the model involved comparing the key process variables with their corresponding standard values. For this, the mathematical model was simulated for 700 s with the initial conditions given in the Supplementary Information. The first 500 s were ignored to let the model approach a *time periodic stationary state* (TPSS) and the state of the system in the remaining time was compared with standard values. The TPSS behavior of a few respiratory and circulatory variables is shown in Figure 3.

At TPSS, the heart rate was 64 beats per minute and the respiratory rate was 11.7 breaths per minute representing normal values for a healthy individual. Figure 3 (a) shows the variations in alveolar volume and the alveolar pressures. The lung volume and alveolar pressure form periodic functions that repeat with each cycle of inhalation and exhalation. The alveolar volume increases when the alveolar pressure is lower than 760 mmHg and vice-versa. Figure 3 (b) shows the oxygen concentrations within the circulatory system. There is a notable difference in the oxygen levels between the pulmonary arteries and veins due to the exchange of gases happening at the lungs and the body tissues. The periodicity due to both heart and respiratory rates is evident in the arterial gas concentrations. The dynamics due to the respiratory cycles are most prominent in the pulmonary veins and these almost vanish by the time the blood reaches pulmonary arteries due to the mixing happening in the ventricles. The faster dynamics due to the cyclic beating of the heart are evident in the oxygen concentrations of pulmonary veins. Figure 3 (c) shows the profiles of blood pressure and flow rates in systemic arteries which are periodic with the heart rate. The blood pressure varies between 70 and 95 mmHg which is close to the normal range of 80-120 mmHg [39]. The mean arterial flow rate of 69ml/s is close to the average value of 83 ml/s mentioned in [39]. Figure 3 (d) shows the model predictions (solid line) for the inflow rate of air in one respiratory cycle. The expiratory flow rate asymptotically approaching zero towards the end of the respiratory cycle is representative of the passive exhalation in humans. On the other hand, the transition from inhalation to exhalation is more rapid due to the abrupt relaxation of the muscles. The profile is also in agreement with the model by Albanese et al. [16], shown by the dashed line.

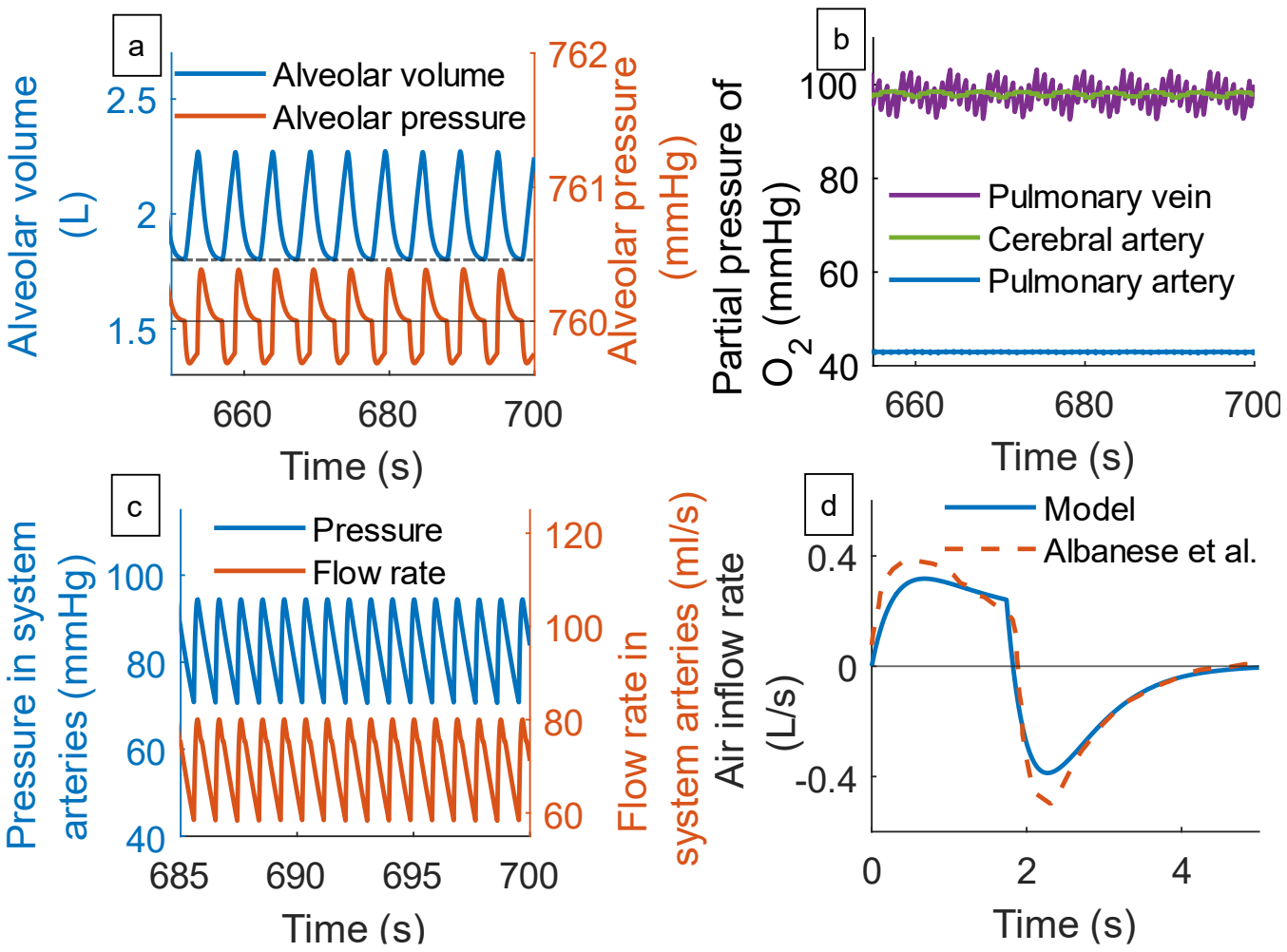

*Figure 3 The dynamics of (a) lung volume and pressure, (b) concentration of oxygen in blood (c) arterial blood flow and pressure, at TPSS. (d) Comparison of air inflow rate in normal breathing with the model given in [16].*

## B. Decreased level of inspired $O_2$ concentration

In the next step, we tested the performance of the proposed cardio-respiratory model under varying environmental conditions, similar to the experiments performed by Dripps et al[43]. It is important to emphasize that we have not tried to fit any parameters in the model to mimic the specific experiment.

Dripps et al.[43], determined the response of the human respiratory and circulatory systems to anoxemia – deficiency of oxygen in the arterial blood – by exposing their subjects to atmospheres with low oxygen concentrations such as commonly observed in airplanes or at high altitudes. In this experiment, the oxygen concentration in inspired air was dropped from normal levels to 10% in a stepwise fashion with respect to time as shown in Figure 4. In the figure, we can see that gas concentrations of not just $O_2$, but also $CO_2$ in the alveoli and the blood progressively decreased. The $CO_2$ behavior is due to the increased ventilation (shown later in Figure 5) that promotes the easy removal of carbon dioxide.

Figure 5 (b&d) shows the TPSS values of changes in heart rate and minute ventilation given by the model. These were obtained by taking the mean of the heart and respiratory rates in the final 200 s at each level of inspired oxygen concentration. Both heart and respiratory rates increased with decreased oxygen levels in inspired air and their gradients were also higher at lower oxygen levels, as observed by Dripps et al [43] (Figure 5, a&c).

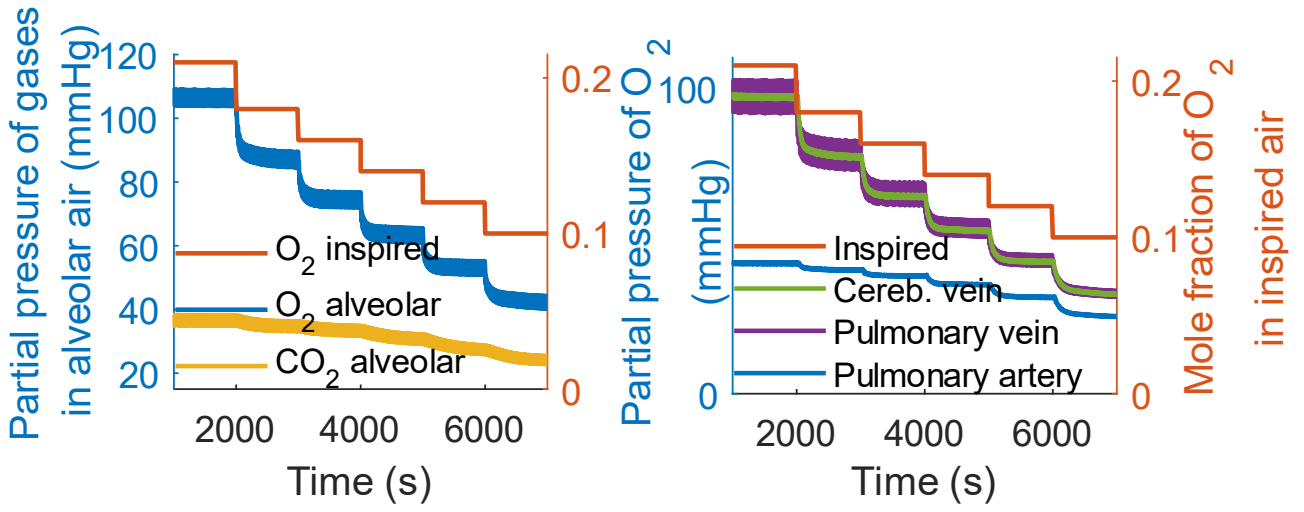

*Figure 4 Changes in (a) alveolar gas concentrations and (b) oxygen in blood, with variations in inspired oxygen concentration*

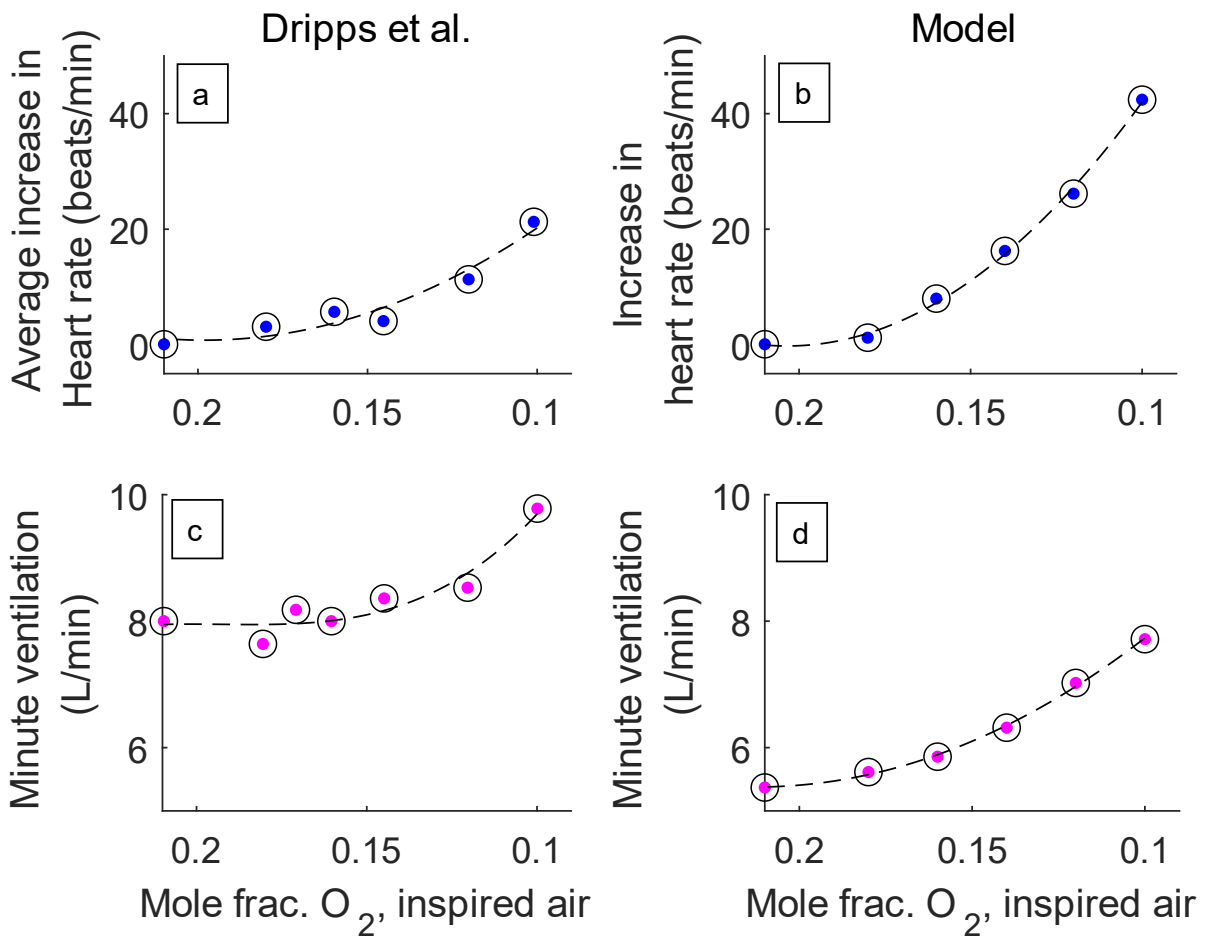

*Figure 5 Variations in heart rate and minute ventilation measured by Dripps et al.[43] (a&c) and obtained from simulations (b&d)*

## C. Replicating a spirometry test

In COPD patients, the obstruction to airflow in the lungs is usually diagnosed using a spirometry test. During this test, the patients are asked to take a deep inhalation and then exhale the air as quickly as they can into a spirometer. The spirometer measures the air capacity in the lungs and how fast the person expires the air, and calculates important respiratory variables such as Forced Vital Capacity (FVC), $FEV1$ ratio (the ratio of the volume of air expired out in 1 s during a forced expiration to that of the total volume of air expired out: $\frac{FEV1}{FVC}$), and Peak Expiratory Flow Rate (PEFR). A significant amount of work has been done on the correlation between these variables and the severity of COPD [44,45].

In the mathematical model, we replicated a spirometry test by modifying the controller output ($R_p$) to simulate a deep breath (see Supplementary Information for details of the implementation). The obtained air inflow rate is given in Figure 6 (a). The time up to $t1$ correspond to the TPSS and a deep inhalation starts at $t_1$. At $t_2$ the air is exhaled out quickly and the expiratory flow rate is plotted against volume in Figure 6 (b, spirogram). The simulation was designed to match the $FVC$ value of 4 L given in [46]. The PEFR obtained is 6.7 L/s, and $\frac{FEV1}{FVC} = 0.83$ which is close to the standard population average[46]. It may also be noted that the terminal part of the spirogram has a slope that is almost constant, another characteristic of healthy lung function [47].

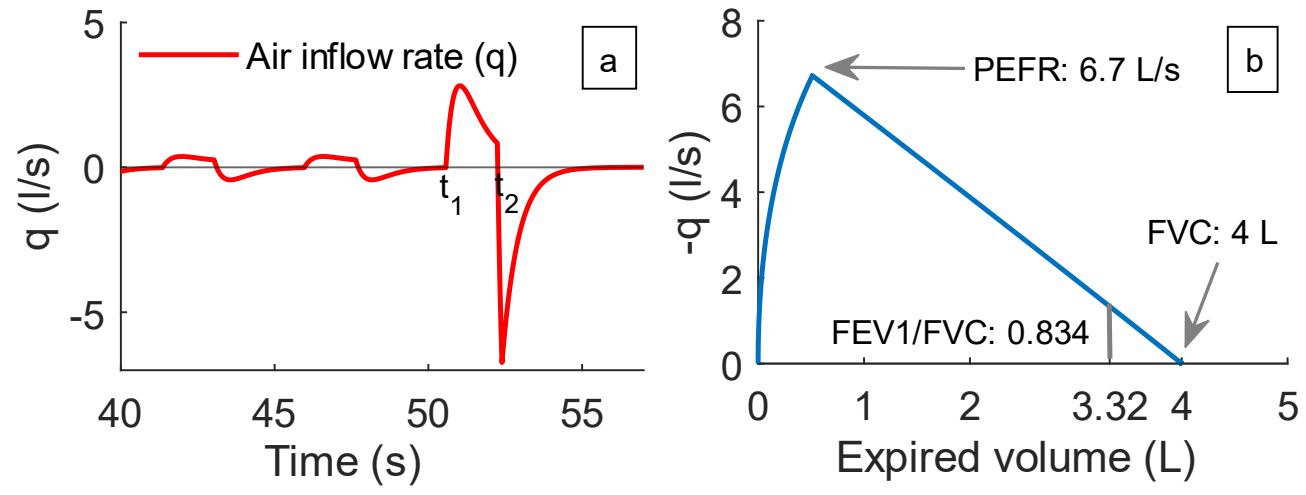

*Figure 6 (a) Changes in air inflow rate during a spirometry. The spirogram is given in (b).*

## IV. Model adaptation to represent a COPD patient

Having verified the model to be qualitatively representative of the normal (healthy) cardio-respiratory system, the next goal is to adapt the model to represent a COPD patient. For this, we identified the components of this model that were associated with the manifestations of COPD. In this section, we explain the procedure followed to identify and perform a quantitative analysis of the COPD-related components of the model.

### *A. Identification of COPD-related components in the cardio-respiratory model*

COPD is a heterogeneous, multi-component disease causing damage to different parts of the cardio-respiratory system. We start by examining the three primary manifestations of COPD namely, (1) airway obstruction, (2) emphysema, and (3) pulmonary hypertension and the components in the systems biology model associated with them.

#### *1)* Airway Obstruction*:*

Airflow in the lungs is the result of the balance between the elastic recoil of the lungs promoting flow and the resistance of the airways that limits flow. In COPD patients, and in particular people with *small airways disease*, excessive mucus (usually as a result of long-term smoking) narrows the bronchi. Consequently, the airway resistance increases, resulting in flow limitation. This COPD symptom can be replicated in simulation by increasing the value of the parameter representing 'Airway Resistance' ($R$ in Eq. S.3), from the 'nominal' value for a healthy subject to a value high enough to change the process variables, $q$ (air flow rate) sufficiently to produce effects matching what occurs in COPD patients.

#### *2) Lung Hyperinflation*

Alveoli in the lungs are separated from one another by the alveolar septum. In *emphysema*, alveoli coalesce due to destruction of the septum, with the following two consequences. First, since the alveolar chambers in coalesced alveoli will now be much larger than the original distinct alveoli, the total surface area per volume of the respiratory membrane decreases. Second, due to the loss of elastic recoil, air sacs in the lungs lose their ability to function and exhale the air properly. As carbon dioxide is trapped in the lungs, fresh air flowing into the lungs pushes the walls of the lung further out with each new breath. This lack of air transfer causes the lungs to expand and lose their elasticity even further. The loss of elasticity causes more carbon dioxide to remain in the lungs leaving less space for fresh air and causing shortness of breath. Over time, the muscles and ribs surrounding the lungs are forced to stretch to fit the over-expanded lungs. The diaphragm, the major muscles used for breathing, becomes flattened and loses its ability to function properly. In the systems biology model, we denote the elasticity of air sacs in the lungs by the parameter $E_T$ (total elastance), and the unloaded lung volume by $V_0$. Changes in $E_T$ and $V_0$, result in substantial changes in physiological variables such as Total Lung Capacity (TLC) and Residual Volume (RV) which are of interest in emphysema.

#### *3) Pulmonary Hypertension (PH):*

PH [9] is a type of high blood pressure that affects the pulmonary arteries and the right side of the heart. PH is a common complication in COPD wherein, pulmonary arterioles and capillaries become narrowed, blocked, or destroyed, making it harder for blood to flow through the lungs, thereby raising pressure within the lungs' arteries. As the pressure builds, the heart's right ventricle must work harder to pump blood through the lungs, eventually causing the heart muscles to weaken and ultimately fail. In the systems biology model, we identified the parameter $R_{pa}$ (pulmonary artery resistance) as another COPD-related parameter, the increase of which causes the corresponding increase in the pulmonary artery pressure ($P_{pa}$) – an indicator of PH [48].

### *B. Quantitative analysis on the COPD-related parameters*

Having identified important parameters of the systems biology model that are linked to COPD manifestations, our next goal is to determine what changes to the associated parameters are required for the manifestation of COPD responses. We start by changing the value of a COPD-related parameter while keeping the other parameters constant. We monitor the changes in the COPD variables that are good indicators of the symptom associated with the varying parameter until we observe a COPD-like response in the variables. The parameter value at which COPD manifestation occurs is representative of the disease state. The application of this approach to each parameter is described in the following sections.

#### *1)* Airway Resistance (***R***)

In this work, so far, the airway resistance ($R$) has been assumed to be a constant. Though in reality, the resistance has a dependence on the volume of the lung (channels become narrower as the lung volume reduces), this effect is minimal above the functional residual capacity (FRC) in healthy individuals [49]. As the lung volume is usually above the FRC (black dotted line in Figure 3a), we could ignore the effect and assume the resistance to be a constant. On the other hand, in COPD patients, not only does the airway resistance increase, but the sensitivity of the airway resistance to lung volume also becomes more pronounced [22]. To capture this behavior, the airway resistance in the model of a COPD patient was defined as a function of the lung volume as given by Equation 1. Here, $\alpha$ is a parameter indicating the severity of the disease and at $\alpha = 0$, the airway resistance becomes a constant (healthy individual). At positive values of $\alpha$, the airway resistance has an inverse relationship with the lung volume, $V_A$.

$$R_{COPD} = R(1 + \alpha(6 - V_A)) \quad (1)$$

To identify the $\alpha$ (and thereby the $R_{COPD}$) value that is diagnostic of COPD, we used a diagnostic criterion that is widely used: $\frac{FEV1}{FVC} < 0.7$ [1]. The 'spirometry test' (see Supplementary Information for details) was repeated for different values of $\alpha$ until the $FEV1$ ratio decreased to 0.7. At $\alpha = 0.205$ – corresponding to an 82% increase in airway resistance at a lung volume of $2L$ – the $FEV1$ ratio reduced to the desired level (Figure 7a). Due to the increase in airway resistance, the PEFR also dropped by about 27%, in accordance with an alternative diagnostic criteria discussed in Jackson et al.[13] (20% decrease in PEFR). The terminal part of the spirogram also developed an inward curvature which is a hallmark of COPD [50]. Figure 7b gives the dynamics of expired volume obtained from the 'spirometry test'. The total expired volume (FVC) is reduced in the case of high airway resistance. Though it has been hypothesized that the reduction in FVC could potentially be a biomarker in COPD, this is yet to be clinically validated[51].

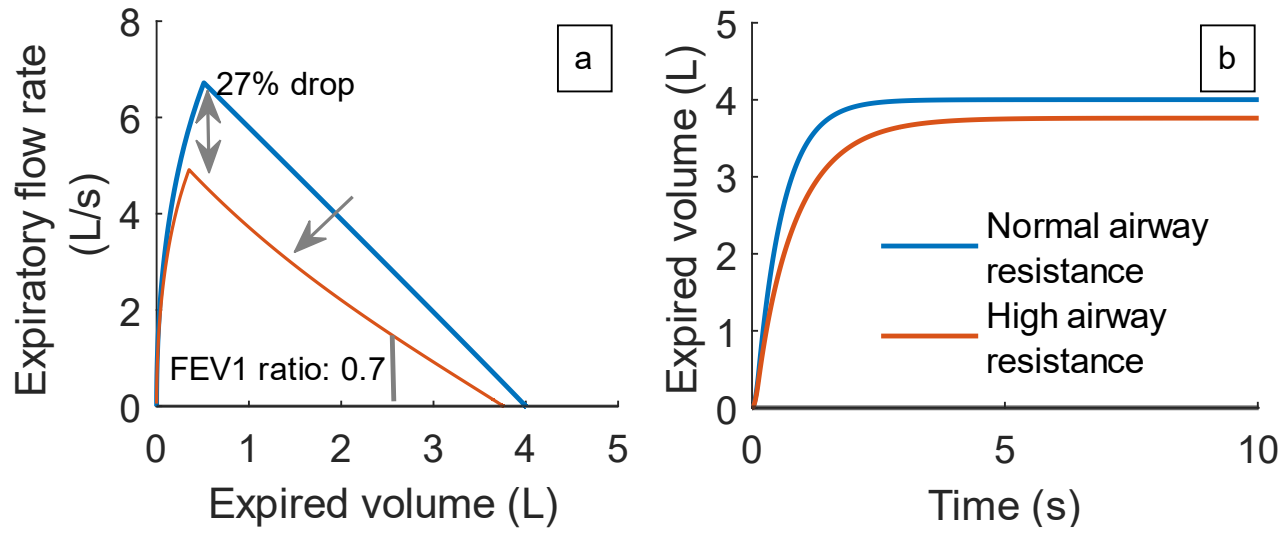

*Figure 7 Spirogram obtained at higher airway resistance. The red lines indicate the profile at high airway resistance ($\alpha = 0.205$).*

### 2) *Chest elastance and unloaded lung volume ($E_T, V_0$)*

COPD patients with emphysema exhibit an increase in their lung volumes and a significant amount of work has been done on recording these changes, e.g., by Biselli et al[52]. According to their results, TLC increases by around 660 ml, and the FRC increases by about 0.9 L (after being scaled by the FRC) in the COPD state compared with the healthy state. We consider these volume changes as the threshold for the particular COPD manifestation.

We simulated the model under the condition of breathing normal air for about 50.5 s to let the system reach TPSS. At $t \cong 50.5\ s$, a deep inhalation was simulated (see Supplementary Information, spirometry test). The FRC was identified as the lowermost point of the normal breathing cycles and the TLC was recorded at the end of inhalation where the lung volume is at its maximum capacity. The procedure was repeated for different values of $E_T$ and $V_0$ until a 900 ml increase in FRC and a 660 ml increase in the TLC were observed (Figure 8a). The solution converged to an unloaded lung volume ($V_{0,COPD}$) of 1.06 L and a chest elastance ($E_{T,COPD}$) of 2.75 mmHg/L.

### 3) *Pulmonary artery resistance ($R_{pa}$)*

In patients with COPD, pulmonary hypertension (PH) is observed due to an increase in $R_{pa}$. PH is defined to be the condition of $P_{pa} > 35$ mmHg [48]. To determine the deviation in $R_{pa}$ from normality necessary to precipitate PH, the model was simulated under the conditions of breathing normal air and a resting metabolic rate for 110 s. The average value of $P_{pa}$ was recorded for the last 10 s. We repeat the simulation for different values of $R_{pa}$ until an average $P_{pa}$ of 35 mmHg was observed. This was observed at $R_{pa,COPD} = 0.3698\ [mmHg.\frac{s}{mL}]$ (originally at 0.198). Figure 8 (b) shows the dynamic response of $P_{pa}$ to changes in $R_{pa}$.

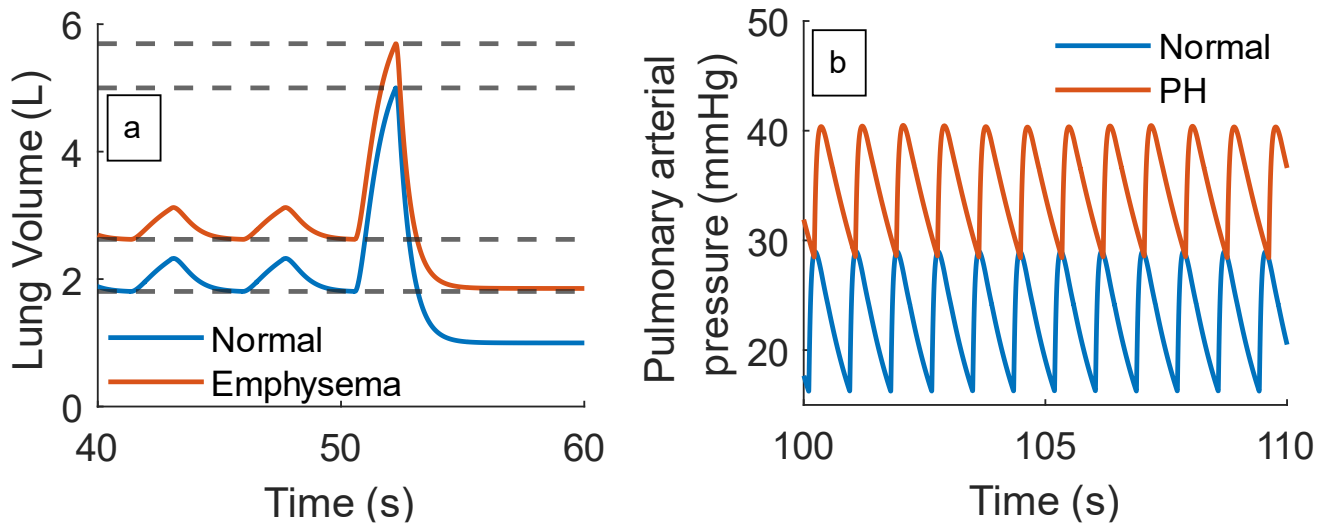

*Figure 8 (a)The changes in lung volume corresponding to emphysema. (b) The pulmonary arterial pressure in PH.*

Table 2 summarizes the parameters chosen to capture the adaptations in COPD, and their values in both healthy and disease states.

*Table 2 Model adaptation for COPD.*

| Condition | Parameter affected | Value: COPD | Value: healthy | Model Eq. |
|---|---|---|---|---|
| Small airways disease | Airway resistance [mmHg s /l] | $R_{COPD} = 1 + 0.205 \times (6 - V_A)$ | $R = 1$ | S.3 |
| Lung hyper-inflation | Unloaded lung volume [l] and Chest elastance [mmHg/l] | $V_{0,COPD} = 1.05$ $E_{T,COPD} = 2.75$ | $V_0 = 0$ $E_T = 2.5$ | S.4 |
| PH | Pulmonary resistance [mmHg.s/ml] | $R_{pa,COPD} = 0.370$ | $R_{pa,COPD} = 0.198$ | S.16 |

## V. Multi-Variable Analysis

The previous analysis is useful for identifying individual manifestations of COPD. However, COPD patients often exhibit a combination of symptoms, meaning changes in different parameters might coexist. In order to provide a comprehensive picture of the effects of COPD on the cardio-respiratory system, we need to develop a holistic approach which takes into account the effects of different parameters and their interactions together on the cardio-respiratory system. This objective can be achieved by proper design of simulation case studies, where the effects of parameter perturbations are studied individually and in combination.

We performed eight different simulations (details in the Supplementary Information) corresponding to combinations of COPD symptoms. The response variables recorded were FVC, TLC, PEFR, $FEV1$ ratio, heart rate, $P_{pa}$, respiratory rate, and minute ventilation. Among these variables, the first four are obtained by simulating deep breath, and are referred to as respiratory characteristics. The last four are the stationary state values obtained from an extended period simulation, referred to as TPSS characteristics.

### A. *Respiratory characteristics*

The FVC, TLC, PEFR and the $FEV1$ ratio were identified for each case by simulating a deep breath (see Supplementary Information, spirometry test). The values observed under each condition are summarized in Figure 9. Changes in both airway resistance and lung elastance led to substantial changes in the vital capacity, while pulmonary arterial resistance had no noticeable effect on it. The total lung capacity increased under emphysema and an increased resistance of the airways led to decrease in both PEFR and the $FEV1$ ratio. On performing the t-tests, the airway resistance had a significant effect on all response variables and the lung elastance showed a significant effect on FVC, TLC and FEV1 ratio (see Table 3). Pulmonary arterial resistance did not have a significant effect on any of the responses considered because it affects the cardio-vascular system primarily and not the respiratory characteristics.

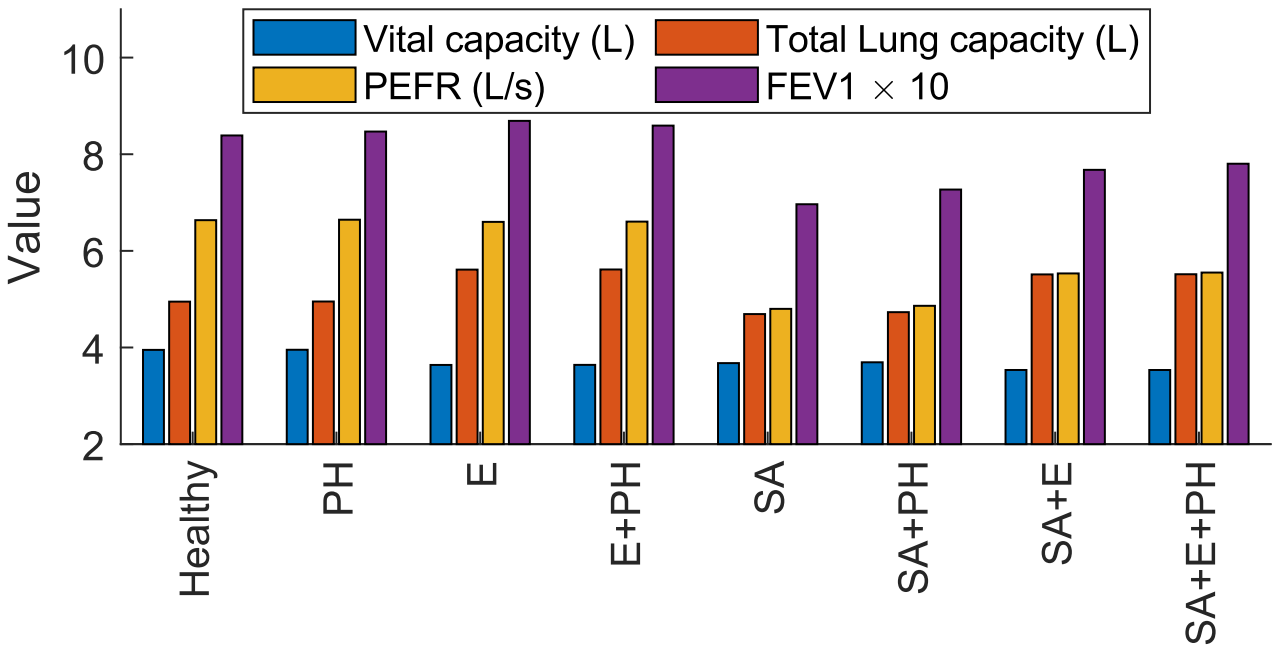

*Figure 9 Respiratory characteristics under varying manifestations of COPD. PH: Pulmonary hypertension, E: Emphysema, SA: Small airways*

*Table 3 The p-values associated with the effect of each factor on the respiratory characteristics. Cells indicating a statistical significance (α= 0.1) are highlighted in green.*

| Factors | Airway resistance | Chest elastance | Pulmonary resistance |
|---|---|---|---|
| FVC | 0.0101 | 0.0046 | 0.9055 |
| TLC | 0.0093 | 0 | 0.7532 |
| PEFR | 0.0016 | 0.1467 | 0.9037 |
| FEV1 | 0.0009 | 0.0288 | 0.4553 |

### B. *TPSS characteristics*

Next, the model was simulated for an extended period, under two different levels of inspired oxygen concentrations – 21% and 12%. Simulations were run for 1000 s under each condition and the reported TPSS values are the mean of the final 100 s of these simulations. The objective of this study was to identify the disease characteristics that could adversely impact the patient under hypoxic environments such as high altitude or while on an airplane.

Figure 10 shows the heart rate, pulmonary artery pressure, respiratory rate and minute ventilation, both in normoxic environments (blue) and hypoxic environments (red). Under normoxic conditions, the changes in the airway resistance and the lung elastance lead to an increased heart rate and all three factors contribute to an increased respiratory rate. Pulmonary resistance stands out as the major contributor to an increase in pulmonary arterial pressure under normoxic conditions. An increase in airway resistance has a major impact on the reduction in minute ventilation due to the difficulty in inhaling and exhaling air. On comparing the *healthy state* with the case of *increased pulmonary resistance* under normoxic conditions, there is a marginal decrease in heart rate (64.2 to 63.8 beats/min) and this is compensated for by a marginal increase in minute ventilation (5.40 to 5.45 L/s). Although the changes are minor, it is still interesting to note how the control systems transfer a part of the effort from the cardiovascular to the respiratory system in presence of a disturbance in the cardiovascular system (the change in pulmonary resistance).

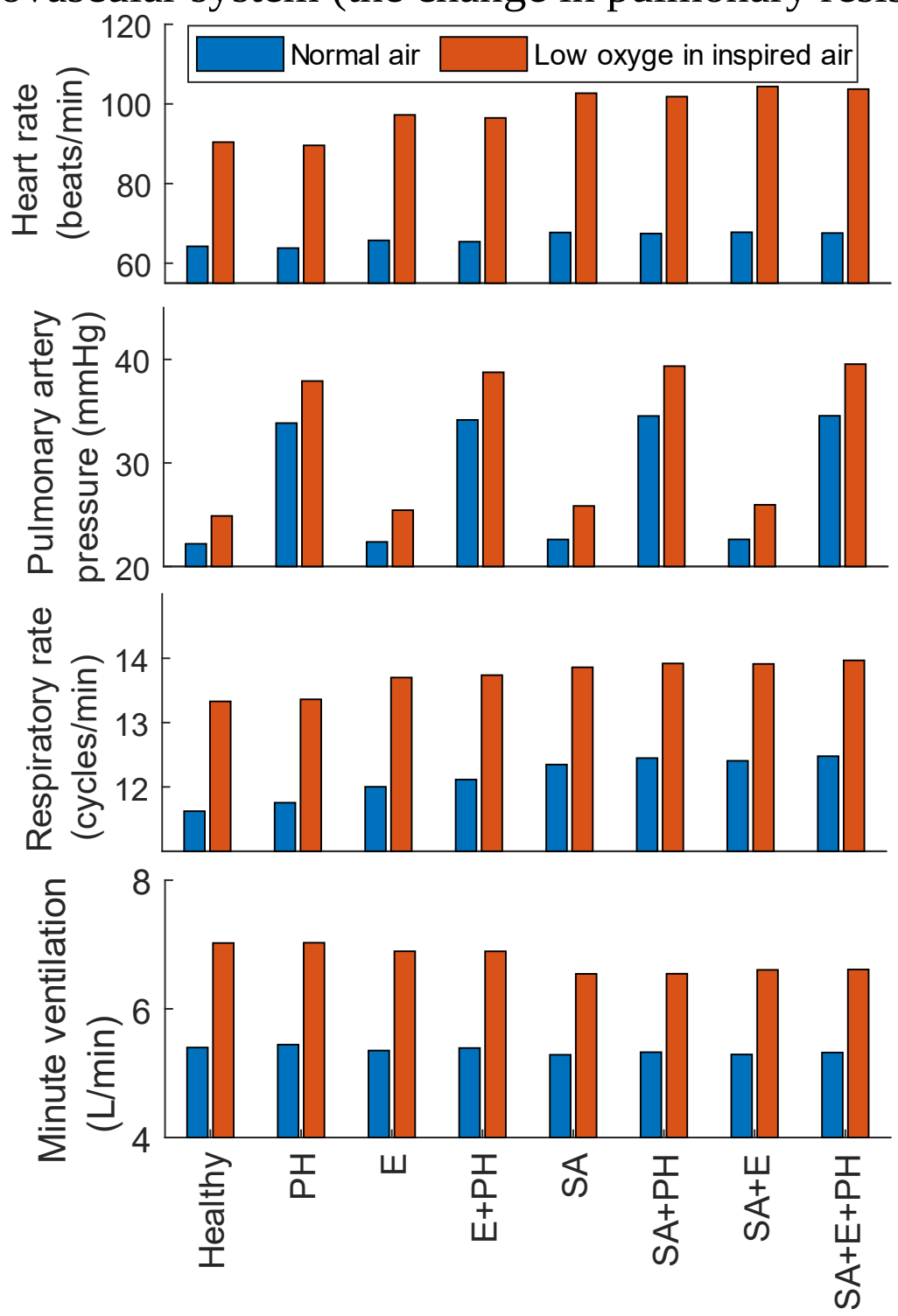

*Figure 10 TPSS characteristics under varying manifestations of COPD. PH: Pulmonary hypertension, E: Emphysema, SA: Small airways*

In the models considered here, under hypoxic conditions, the TPSS characteristics remained at a higher level compared with those at normoxic conditions. To identify the COPD characteristics that have a significant impact on the *changes* in these response variables, t-tests were performed. Airway resistance was identified to have a significant impact on the changes in every response variable considered here (ref. Table 4). That is, when moving from a normoxic environment to a hypoxic environment, the changes observed in all four TPSS characteristics of a person with high airway resistance would be *significantly* different from that of a person with normal airway resistance. Likewise, the changes in heart rate and pulmonary pressure in a person with emphysema would be significantly different from that of someone with normal lung elastance. Pulmonary resistance has significant effects on the changes in pulmonary arterial pressure and respiratory rates.

With all effects being significant, patients with high airway resistance appear to be at the highest risk from changes in atmospheric oxygen levels. Changes in pulmonary arterial pressures are affected by every factor considered here indicating that a decrease in the level of oxygen in air may induce higher stress on the cardiovascular system of all patients, irrespective of the expressed COPD manifestations.

*Table 4 The p-values associated with the effect of each factor on the TPSS characteristics. Green highlight indicates statistical significance (α= 0.1).*

| **Factors →<br>↓Change in:** | **Airway resistance** | **Chest elastance** | **Pulmonary resistance** |
|---|---|---|---|
| Heart rate | 0.0016 | 0.0182 | 0.6334 |
| Pulmonary pressure | 0.0067 | 0.0352 | 0.0001 |
| Respiratory rate | 0.0005 | 0.7693 | 0.0223 |
| Minute ventilation | 0.0013 | 0.8351 | 0.3852 |

## VI. Conclusions

Based on the premise that the regulation of physiological processes is achieved by biological control systems, we developed a control engineering framework for the modeling and analysis of COPD. Specifically we: 1) Developed a representation of the human cardiorespiratory system in terms of control engineering components; 2) Developed a computational dynamic model for each block with differential and algebraic equations representing known physiological functions; 3) Verified the cardio-respiratory model with standard/literature data via simulations; 4) Identified the model parameters associated with physiological characteristics of COPD and quantified the changes in them that are diagnostic of disease state; and 5) Employed a statistical design of experiment approach to investigate the effects these parametric changes exert on the cardiorespiratory system under normal and hypoxic environments.

Our approach facilitated understanding underlying mechanisms of COPD, as it views COPD appropriately as a malfunction (or failure) of one or more components of the overall system. Among the COPD related parameters studied here, an increase in airway resistance is identified to be the disease manifestation with maximum impact, affecting almost all response variables considered in this study. From the model simulations, it is also inferred that moving from normoxic to hypoxic environments induces a significantly higher stress on the cardiovascular system in COPD patients, irrespective of the manifestations observed in them.

# SUPPLEMENTARY INFORMATION

## 1. A CONTROL ENGINEERING MODEL OF THE CARDIO-RESPIRATORY SYSTEM

Here we describe the model equations used to represent different physiological processes in the human cardio-respiratory system. (Figure 1 & 2).

### *1.1. Process: Lungs and Cardiovascular System Compartments*

There are three main processes involved in the lungs and heart to deliver oxygen to and eliminate carbon dioxide from the body cells. Ventilation, in which the inhalation and exhalation occur, diffusion, which is the spontaneous movement of molecules between the gas in the alveoli and the bloodstream in the lung capillaries, and perfusion, the process by which the heart circulates blood throughout the cardiovascular system[53]. We describe the equations used for modeling each of these in the following paragraphs.

#### *1.1.1. Ventilation*

It is assumed that the lung is a single compressible container which can support gas exchanges via blood flowing to and from the pulmonary capillaries. The flow of air in and out of the lungs $(q)$ is driven by the difference between the mouth pressure $(P_m)$ and the alveolar pressure $(P_A)$. Among these, the mouth pressure remains constant, and the alveolar pressure varies with time, depending on the total inflow of gases into the lung $(Q_A)$ and the pleural pressure $(P_L)$. Figure 11 gives a schematic describing these processes and the mathematical relations are summarized by Equations S.1 – S.4. These equations are based on the works of Ben-Tal and others[32,33].

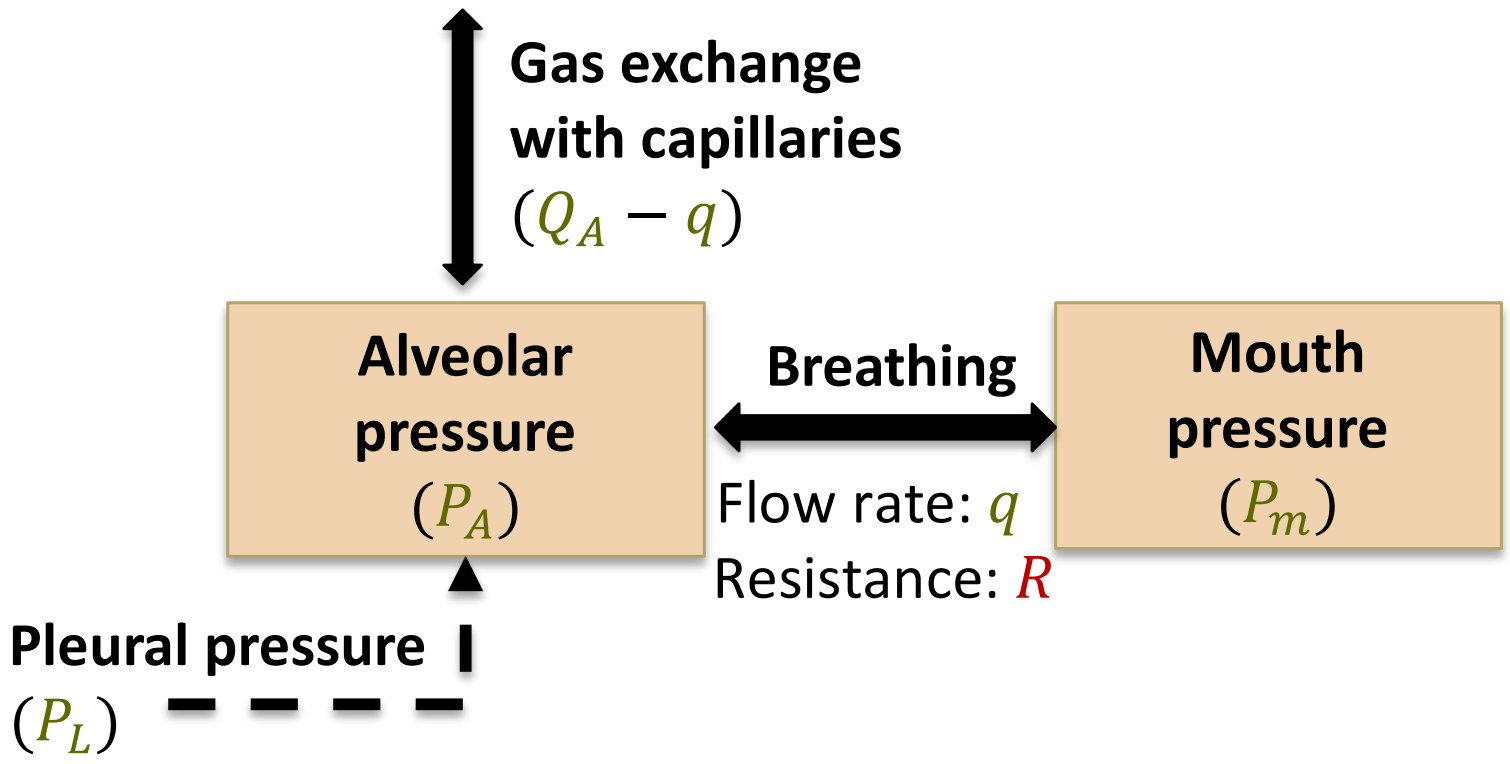

*Figure 11 A schematic of the blocks involved in ventilation.*

$$\frac{dP_A}{dt} = \frac{P_m E_T}{P_A} Q_A + \frac{dP_L}{dt} \tag{S.1}$$

$$Q_A = q + D_{CO2}(p_{CO2,Sa} - p_{CO2,al}) + D_{O2}(p_{O2,Sa} - p_{O2,al}) \tag{S.2}$$

$$q = \frac{P_m - P_A}{R} \tag{S.3}$$

$$V_A = \frac{P_A - P_L}{E_T} + V_0 \tag{S.4}$$

#### *1.1.2. Diffusion*

In the alveoli, oxygen and carbon dioxide are exchanged between the alveolar air and pulmonary capillaries through diffusion as shown in Figure 12. Ben-Tal[32,33] assumed the volume of the lung capillaries to be same as the stroke volume, allowing the blood to remain at the capillaries for the duration of a heartbeat undergoing gas exchange. This allowed them to reinitialize the differential equations at every heartbeat to account for blood circulation. In the present work, we modified their model, and integrated it with the model of the cardiovascular system[34] to account for the blood circulation in the pulmonary capillaries.

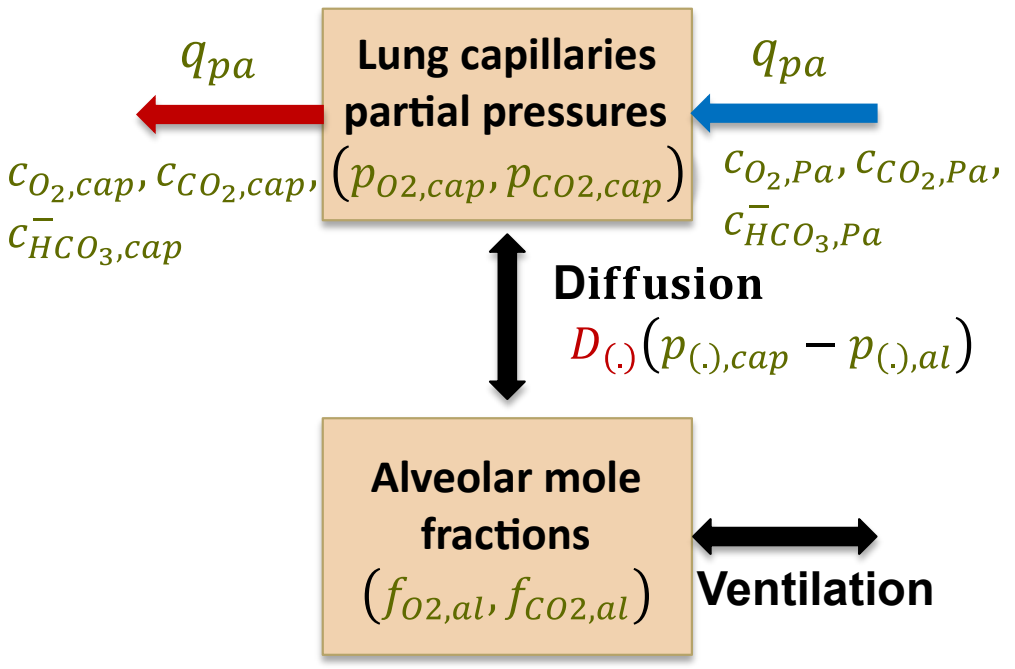

*Figure 12 Schematic showing the key variables involved in the diffusion of gases between alveolar air and the lung capillaries.*

The relations used to model the diffusion and the subsequent changes in gas concentrations are given by Equations S.5 – S.15. Equations S.5 and S.6 describe the change in mole fractions of oxygen $(f_{O2,al})$ and carbon dioxide $(f_{CO2,al})$ in the alveolar air due to the diffusion and ventilation. The alveolar partial pressures of gases $(p_{(.),al})$ are calculated from the mole fraction of gases and the alveolar pressure, after subtracting the partial pressure of water vapor $p_w$. The inspired volume $V_i$ represents the amount of air inhaled in the current respiratory cycle and is brought to zero during exhalation as defined by Equation S.9. The mole fraction of oxygen in the inspired air $(f_{O2,i})$ is assumed to be the same as the concentration of oxygen in the lung dead space in the early phase of inhalation. Later on, after all air in the dead space is inhaled, it becomes equal to the mole fraction of atmospheric air. Finally, during exhalation, $f_{O2,i}$ is made equal to the mole fraction of alveolar oxygen (Equation S.10). Equations S.11 and S.14 model the changes in partial pressure of oxygen $(p_{O2,cap})$ and carbon dioxide $(p_{CO2,cap})$ in the lung capillaries due to diffusion and circulation of blood. The integration of the terms for circulation into these equations is based on the work by Ellwein et al[34]. The functions $\tilde{f}$ and $g$ are the saturation functions for oxygen in blood[33,34]. The last equation (Equation S.15) models the change in bicarbonate ions in the blood.

$$\frac{df_{O2,al}}{dt} = \frac{1}{V_A}\Big[D_{O2}(p_{O2,cap} - p_{O2,al}) + q(f_{O2,i} - f_{O2,al}) - f_{O2,al}\Big(D_{CO2}(p_{CO2,cap} - p_{CO2,al}) + D_{O2}(p_{O2,cap} - p_{O2,al})\Big)\Big] \quad \text{(S.5)}$$

$$\frac{df_{CO2,al}}{dt} = \frac{1}{V_A}\big[D_{CO2}(p_{CO2,cap} - p_{CO2,al}) + q(f_{CO2,i} - f_{CO2,al}) - f_{CO2,al}(D_{O2}(p_{O2,cap} - p_{O2,al}) + D_{CO2}(p_{CO2,cap} - p_{CO2,al}))\big] \quad \text{(S.6)}$$

$$p_{O2,al} = f_{O2,al}(P_A - p_w) \quad \text{(S.7)}$$

$$p_{CO2,al} = f_{CO2,al}(P_A - p_w) \quad \text{(S.8)}$$

$$V_i(t) = \begin{cases} \dfrac{(2P_m - P_A(t) - P_A(t-\Delta t))}{2R}\Delta t + V_i(t-\Delta t) & if \quad (P_m - P_A) \geq 0 \\ 0 \quad if \quad (P_m - P_A) < 0 \end{cases} \quad \text{(S.9)}$$

$$f_{O2,i} = \begin{cases} f_{O2,m} & if \quad V_i > V_D \quad and \quad (P_m - P_A) > 0 \\ f_{O2,d} & if \quad V_i \leq V_D \quad and \quad (P_m - P_A) > 0 \\ f_{O2,al} & if \quad (P_m - P_A) \leq 0 \end{cases} \quad \text{(S.10)}$$

$$\frac{dp_{O2,cap}}{dt} = \frac{D_{O2}}{V_{cap}\,\sigma_{O2}}\left(1 + \frac{4T_h}{\sigma_{O2}}\frac{d\tilde{f}(p_{O2,Pa})}{dp_{O2,Pa}}\right)^{-1}(p_{O2,al} - p_{O2,cap}) + \frac{q_{pa}}{V_{cap}g(p_{O2,cap})}c_{O2,cap} - \frac{q_{pa}}{V_{cap}g(p_{O2,Pa})}c_{O2,Pa} \quad \text{(S.11)}$$

$$\tilde{f}(p_{O2,cap}) = \frac{LK_T\sigma p_{O2,cap}(1 + K_T\sigma p_{O2,cap})^3 + K_R\sigma p_{O2,cap}(1 + K_R\sigma p_{O2,cap})^3}{\left(L(1 + K_T\sigma p_{O2,cap})^4 + (1 + K_R\sigma p_{O2,cap})^4\right)} \quad \text{(S.12)}$$

$$\frac{dg(p_{O2,cap})}{dp_{O2,cap}} = 2K_{O2}k_{O2}e^{-k_{O2}p_{O2,cap}}(1 - e^{-k_{O2}p_{O2,cap}}) \quad \text{(S.13)}$$

$$\frac{dp_{CO2,cap}}{dt} = \frac{D_{CO2}}{\sigma_{CO2}V_{cap}}(p_{CO2,al} - p_{CO2,cap}) + \frac{\delta l_2}{\sigma_{CO2}}hc^-_{HCO3,cap} - \delta r_2 p_{CO2,cap} + \frac{q_{pa}}{V_{cap}}p_{CO2,Pa} - \frac{q_{pa}}{V_{cap}}p_{CO2,cap} \quad \text{(S.14)}$$

$$\frac{dc^-_{HCO3,cap}}{dt} = \delta r_2\sigma_{CO2}p_{CO2,cap} - \delta l_2 hc^-_{HCO3,cap} + \frac{q_{pa}}{V_{cap}}(c^-_{HCO3,cap} - c^-_{HCO3,Pa}) \quad \text{(S.15)}$$

### *1.1.3. Circulation*

As shown in Figure 13, the cardiovascular system is represented by a closed circuit with three arterial compartments, three venous compartments, and two ventricular compartments. These components are coupled with the lungs and the tissue compartments, which are responsible for the exchange of oxygen and carbon dioxide. Each cardiovascular compartment yields a representative transmural pressure $P(t)$, volume $V(t)$ and a constant compliance $C$. Compartments are separated by resistors $R$, analogous to an RC-circuit.

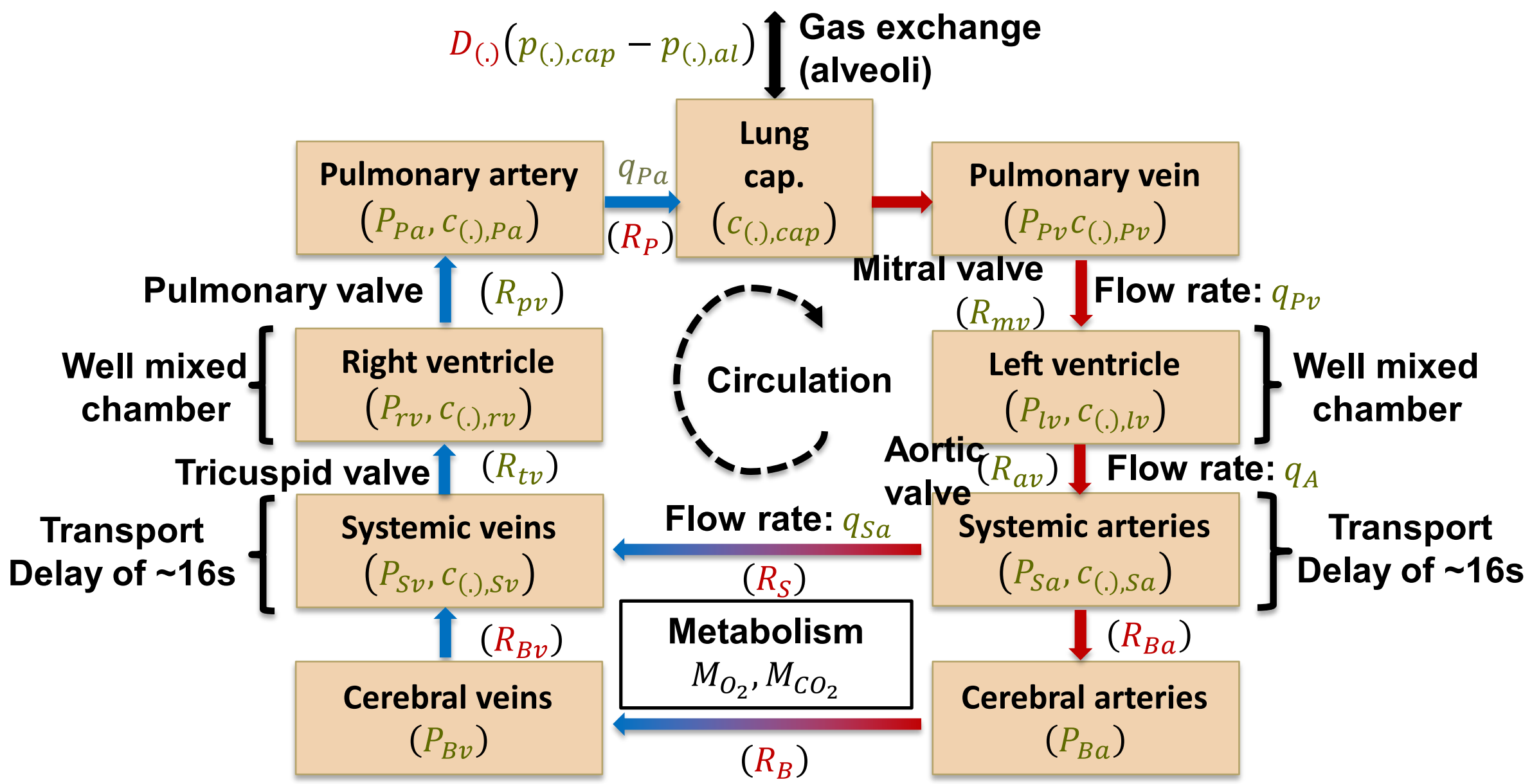

*Figure 13 Schematic showing the key variables involved in the perfusion process. The R's denote the resitances, the q's denote the blood flow rate, and the c's denote the blood gas concentrations. The M's denote the metabolic rates (input variables). The resistances offered by the valves in the heart are functions of the pressures in adjacent compartments and therefore, vary with time.*

As described by Ellwein et al[34]., in each compartment, the dynamics of blood pressure and flow can be explained by the mass and momentum balance. For example, Equation S.16 defines the resistive flow of blood between the compartments - the *pulmonary vein* and the *left ventricle* as a function of the pressures in the respective compartments and the resistance offered by the mitral valve. Similar equations define the flow rates between each pair of compartments in the circulatory system. Among the resistances, the arterial and venous resistances are parameters that remain constant for the entire duration of a heartbeat. However, the resistances at the four valves of the heart (mitral, aortic, tricuspid, and pulmonary) have dynamics that mimic the opening and closing of valves. Equation S.17 describes the variation in resistance across the mitral valve depending on the pressures in pulmonary vein and left ventricle. The equations describing the other three valves are the same, except for the parameter values which are given in Section 3. The variation in pressure in each compartment is a function of the net blood flow into the compartment and the compliance associated with the compartment, as given by Equation S.18. It is assumed that the compliance of the compartments remains constant in time and the values used in this work are available in Section 4. The time varying profile of the ventricular elastances drive the flow of blood in the RC circuit through changes in the ventricular pressures, as defined by Equation S.19. Here $P_{lv}$ is the left ventricular pressure, $E_{lv}$ is the time varying elastance, $V_{lv}$ is the ventricular volume, and $V_{lv,d}$ is the dead volume of the respective ventricle (similarly for the right ventricle).

$$q_{pv} = \frac{P_{pv} - P_{lv}}{R_{mv}} \tag{S.16}$$

$$R_{mv} = min\left[R_{mv,o} + e^{-2(p_{Pv} - p_{lv})}, R_{mv,c}\right] \tag{S.17}$$

$$\frac{dP_{Pv}}{dt} = (q_{Pa} - q_{Pv})/C_{Pv} \tag{S.18}$$

$$P_{lv}(t) = E_{lv}(t)\left[V_{lv}(t) - V_{lv,d}\right] \tag{S.19}$$

The variations in concentrations of blood gases in the compartments of the circulatory system is defined using standard material balance equations. Equations S.11, S.14 (lung capillaries), and S.20, S.21 (tissues) represent the exchange of blood gases between the circulatory system and the rest of the human body. In all other compartments, the difference between the convective transport in and out of the compartment equals the accumulation of gases inside the compartment. The left and right ventricles are represented as well mixed chambers of varying volume as defined by Equation S.22.

$$q_{sa}\left(c_{CO2,Sa} - c_{CO2,sv}\right) = \left(M_{S,CO2}\right) \tag{S.20}$$

$$q_{sa}\left(c_{O2,Sa} - c_{O2,Sv}\right) = \left(M_{S,O2}\right) \tag{S.21}$$

$$\frac{dc_{(.),lv}}{dt} = \frac{q_{Pv}\left(c_{(.),Pv} - c_{(.),A}\right)}{V_{lv}} \tag{S.22}$$

Due to the time required for circulating the blood in the human system, there is a delay for the oxygen rich blood from the lung to reach the tissues. Similarly for carbon dioxide rich blood to reach from tissues to the lung. In the mathematical model, these are incorporated as variable transport delays in two compartments - systemic arteries and systemic veins. The transport delay ($t_d$) at time $t$ in the systemic arteries is calculated by solving Equation S.23 (see [36]), where $\frac{1500}{q_A}$ is the *instantaneous delay* corresponding to the flow rate at time $t$. This gives an average delay ~16 s at resting metabolic rate, which is close to the values reported in [37]. Similarly for the systemic veins. There is zero delay in other compartments.

$$\int_{t-t_d(t)}^{t} \frac{q_A(\tau)}{1500} d\tau = 1 \qquad \text{(S.23)}$$

*1.2. Sensor: Chemoreceptors*

The afferent input into the respiratory control system is provided primarily by two groups of neural receptors: 1) peripheral arterial chemoreceptors; 2) central (brainstem) chemoreceptors[26]. The peripheral arterial chemoreceptors consist of the carotid and aortic bodies. The carotid bodies respond to the partial pressures of oxygen and carbon dioxide and the hydrogen ion concentration in the blood. The intensity of the response varies according to the severity of the arterial hypoxemia or acidosis. The central chemoreceptors respond primarily to alterations of hydrogen ion concentration in the cerebrospinal fluid (CSF) and medullary interstitial fluid. The contribution of these central chemoreceptors to ventilation depends on the factors that alter hydrogen flux in their vicinity, causing changes in their intracellular pH.

In the model, the process of sensing blood gas level and blood pH is represented by three linear static relationships (Equations S.24 – S.26). The first two equations show the relationship between the blood partial pressure of oxygen and carbon dioxide in the systemic arteries and in the brain, and their corresponding measurements ($p_{O_2,s}, p_{CO_2,s}$). The third equation relates the concentration of bicarbonate (as an indicator of blood pH) to the sensor output. An additive relation between the measurements made by the peripheral and central chemoreceptors are used, as suggested by Batzel et al[38]. The constant of proportionality, $K_{(.),s}$, is the sensor gain, which is assumed to be 1 for a healthy person.

$$p_{O_2,s} = K_{O_2,s} \frac{(p_{o2,sa} + p_{o2,B})}{2} \qquad \text{(S.24)}$$

$$p_{CO_2,s} = K_{CO_2,s} \frac{(p_{Co2,sa} + p_{Co2,B})}{2} \qquad \text{(S.25)}$$

$$c_{HCO3,s}^{-} = K_{HCO3^-,s} \frac{(c_{HCO3,sa}^{-} + c_{HCO3,B}^{-})}{2} \qquad \text{(S.26)}$$

*1.3. Controller: Respiratory Control Center*

Breathing rhythm in mammals is controlled by a neural central pattern generator (CPG) that consists of multiple groups of neurons in the pons and medulla[54,55]. These groups of neurons are classified according to their behavior during the different phases of the breathing cycle; i.e., early inspiration, inspiration, and expiration. There is an extensive amount of work on CPG models which tend to focus on the mechanisms of individual neurons during the three phases of breathing cycle[33,56,57]. These models are based on the theory by Hodgkin-Huxley who developed the first quantitative model of the initiation and propagation of an electrical signal along a squid giant axon.

Here we used the model by Ben-Tal[33] for capturing the response of the respiratory control center. The mechanisms behind the generation of the controller response, $R_p(t)$, from the controller input, $p_{o_2,s}$ and $p_{co_2,s}$, can be explained by the two interacting components: 1) An oscillator which receives the signals from the respiratory sensors, $p_{o_2,s}$ and $p_{co_2,s}$, and generate the average measure of spike rate ($A$) in the neurons (Equations S.27 – S.34); 2) An inspiratory pattern generator that receives the signal from the oscillator and transforms it into a ramp pattern of neural activity - $R_p(t)$ (Equations S.35 – S.36).

In the model of the oscillator, $\tilde{g}_t$ is a parameter affecting the conductivity of the sodium ions based on the changes in the blood gas concentrations and $K_{ctrl}$ is a parameter of the inspiratory pattern generator responsible for the respiratory amplitude. Ben-Tal[33] tested Proportional-Integral (PI) functions to relate $\tilde{g}_t$ and $K_{ctrl}$ to the respiratory sensors' input. The exact feedback mechanism in the physiological system is not known yet, but $\tilde{g}_t$ and $K_{ctrl}$ can be calculated from the input-output data available on respiratory control. That is, the response of the respiratory controller (respiratory rate and minute ventilation) to different values of $p_{CO_2,Sa}$ and $p_{O_2,Sa}$ are available and these can be used to tune the controller parameters $\tilde{g}_t$ and $K_{ctrl}$. Plots showing the variations in $\tilde{g}_t$ and $K_{ctrl}$, and alveolar ventilation in response to variations in blood gas levels are provided in Section 6. These are based on the data given by Guyton and Hall[39].

$$\frac{dA}{dt} = \alpha(1 - A) - \beta A + \gamma \qquad \text{(S.27)}$$

$$\alpha = \tilde{g}_{nap}\overline{m}_p h_p \qquad \text{(S.28)}$$

$$\beta = \tilde{g}_t + \tilde{g}_L \qquad \text{(S.29)}$$

$$\gamma = \tilde{b}\tilde{g}_t + \tilde{E}_L\tilde{g}_L \qquad \text{(S.30)}$$

$$\bar{m}_p = \frac{1}{1 + e^{(A-\tilde{\theta}_{mp})/\tilde{\sigma}_{mp}}} \tag{S.31}$$

$$\frac{dh_p}{dt} = \alpha_{hp}\left(1 - h_p\right) - \beta_{hp}h_p \tag{S.32}$$

$$\alpha_{hp} = \frac{1}{2\bar{\tau}_{hp}} e^{-(A-\tilde{\theta}_{hp})/2\tilde{\sigma}_{hp}} \tag{S.33}$$

$$\beta_{hp} = \frac{1}{2\bar{\tau}_{hp}} e^{(A-\tilde{\theta}_{hp})/2\tilde{\sigma}_{hp}} \tag{S.34}$$

$$R_p(t) = \begin{cases} K_{ctrl} & if\ A(t) > T_{r1}\ and\ R_p(t-\Delta t) < T_{r4}, \\ I_p & if\ A(t) > T_{r1} \\ 0 & if\ A(t) \le T_{r2} \\ R_p(t-\Delta t) & if\ \ R_p(t-\Delta t) > T_{r3}\ and\ \left|\dfrac{A(t) - A(t-\Delta t)}{\Delta t}\right| < \varepsilon \end{cases} \tag{S.35}$$

$$I_p = I_l * \left(A(t-\Delta t) + A(t)\right) * \frac{\Delta t}{2} + R_p(t-\Delta t) \tag{S.36}$$

### 1.4. Actuator: Lung Muscles Compartment

The lung is contracted and expanded by two types of muscle: 1) The diaphragm (abdominal muscles) which can have an upward or downward movement to shorten or lengthen the chest cavity; and 2) The external intercostal muscles which can elevate or depress the ribs to increase and decrease the anteroposterior diameter of the chest cavity. The cells in the abdominal and intercostal muscles translate the electrical signal that comes from the controller into mechanical contraction and determines the pleural pressure in the lungs. Here the muscles are considered a spring which can be excited by the controller signal. The dynamics of muscle displacement are explained by Equations S.37 and S.38[33]. The controller feedback is now connected to the process compartment through the pleural pressure $(P_L)$ (see Equation S.1).

$$\frac{dx_m}{dt} = -k_1 x_m + k_2 R_p(t) \tag{S.37}$$

$$P_L = P_m - P_{L0} - k_p x_m \tag{S.38}$$

### 1.5. Controller: Heart Rate Control

As the primary focus of our work is the simulation of the respiratory system, the heart rate was explicitly computed from the blood gas concentrations (Equation S.39) rather than using a detailed model as in the case of respiratory control. This equation was taken from the work of Milhorn et al.[40], who found the relation between the cerebral blood flow and blood partial pressure of oxygen and carbon dioxide. We expanded this equation further to account for the total cardiac output and the heart rate by adding two adjustable parameters, i.e., the ratio of cerebral blood flow to the total cardiac output $(CBF_f)$ and the amount of blood which is pumped at each heartbeat $(H_c)$. The heart rate is then used to calculate the heart period $(T)$, the systolic time interval $(T_M)$ and the diastolic relaxation time $(T_R)$ (Equation S.40).

$$H = \frac{1}{CBF_f * H_c}\left(750 + W\left(h_1\left(P_{CO2,Sa}\right)^5 + i_1\left(P_{CO2,Sa}\right)^4 + j_1\left(P_{CO2,Sa}\right)^3 + p_1(P_{CO2,Sa})^2 + q_1 P_{CO2,Sa} + r_1 + f\left(g - P_{O2,Sa}\right)^s\right)\right) \tag{S.39}$$

$$T = \frac{1}{H}\ [\mathrm{s}]$$
$$T_M = T_{M,f}.T \tag{S.40}$$
$$T_R = T_{R,f}.T$$

### 1.6. Actuator: Heart muscles

The time-varying elastance of the heart muscles in the left and right ventricles is given by Equation S.41. The systolic and early diastolic phases of the cycle are represented as scaled trigonometric functions. The elastance is assumed to remain constant in the rest of the diastole.

$$E(t) = \begin{cases} (E_s - E_D)\left[1 - \cos\left(\dfrac{\pi t}{T_M}\right)\right]/2 + E_D & 0 \le t \le T_M \\ (E_s - E_D)\left[\cos\left(\dfrac{\pi(t - T_M)}{T_R}\right) + 1\right]/2 + E_D & T_M \le t \le T_M + T_R \\ E_D & T_M + T_R \le t \le T \end{cases} \tag{S.41}$$

### 1.7. Actuator: Vasodilation

When a higher flow rate of blood is required in the human body, the blood vessels dilate to reduce the net resistance of the circulatory system and this phenomenon is termed vasodilation. To represent this, in the mathematical model, the resistance of the arteries was divided by factor $(1 \le \tau \le 2.3)$ depending on the heart rate as given by Equation S.42. The relation was developed by fitting a sigmoidal function on existing data[41,42].

$$\tau = \frac{1 + 2.3\, e^{\frac{HR-120}{2.3}}}{1 + e^{\frac{HR-120}{2.3}}} \tag{S.42}$$

## 2. Replicating a spirometry test

In this section, we describe the steps followed in replicating a spirometry test in the mathematical model of a healthy individual. We replicated a spirometry test by manually manipulating the controller's ramp output, $R_p$ (Equation S.37). $R_p$ – the output of the controller – is an electrical neural signal that determines the contraction of lung muscle. When $R_p$ is greater than 0 ($R_p$> 0) inhalation occurs and $R_p$ equals 0 ($R_p = 0$) corresponds to exhalation during the natural breathing rhythm. We started by simulating the model under the condition of breathing normal air and resting metabolic rate. The simulation continued for almost 50.5 seconds until we reached a stationary state. At $t \cong 50.5\ s$, around the end of expiration in a respiratory cycle, a step change in the $R_p$ signal was applied, increasing it by a value of six (Figure 14 a). Higher $R_p$ values resulted in a higher muscle displacement ($x_m$), lower pleural pressure ($P_L$), and thereby, deeper inhalation (Figure 14 b & c). Following the deep inhalation at $t \cong 52.5s$, we quickly reduced the $R_p$ signal to a value of -20 for a short period of time. The purpose of this step was to rapidly bring down the muscle displacement – the model equivalent of the quick release of air at the end of a deep inhalation. Once the muscle displacement reached zero, $R_p$ was increased to a higher (but negative) value of -1.6 and remained there until the end of the simulation. When $R_p$ remains below zero, muscle displacement also turns negative, and a forced exhalation occurs. The specific changes required $R_p$ to simulate the deep breath – the increase by 6 units during the inhalation, the decrease to -1.6 during the exhalation – were identified by solving for the perturbations that result in an FVC of 4 L (TLC of 5L and RV of 1L, Figure 14)[46].

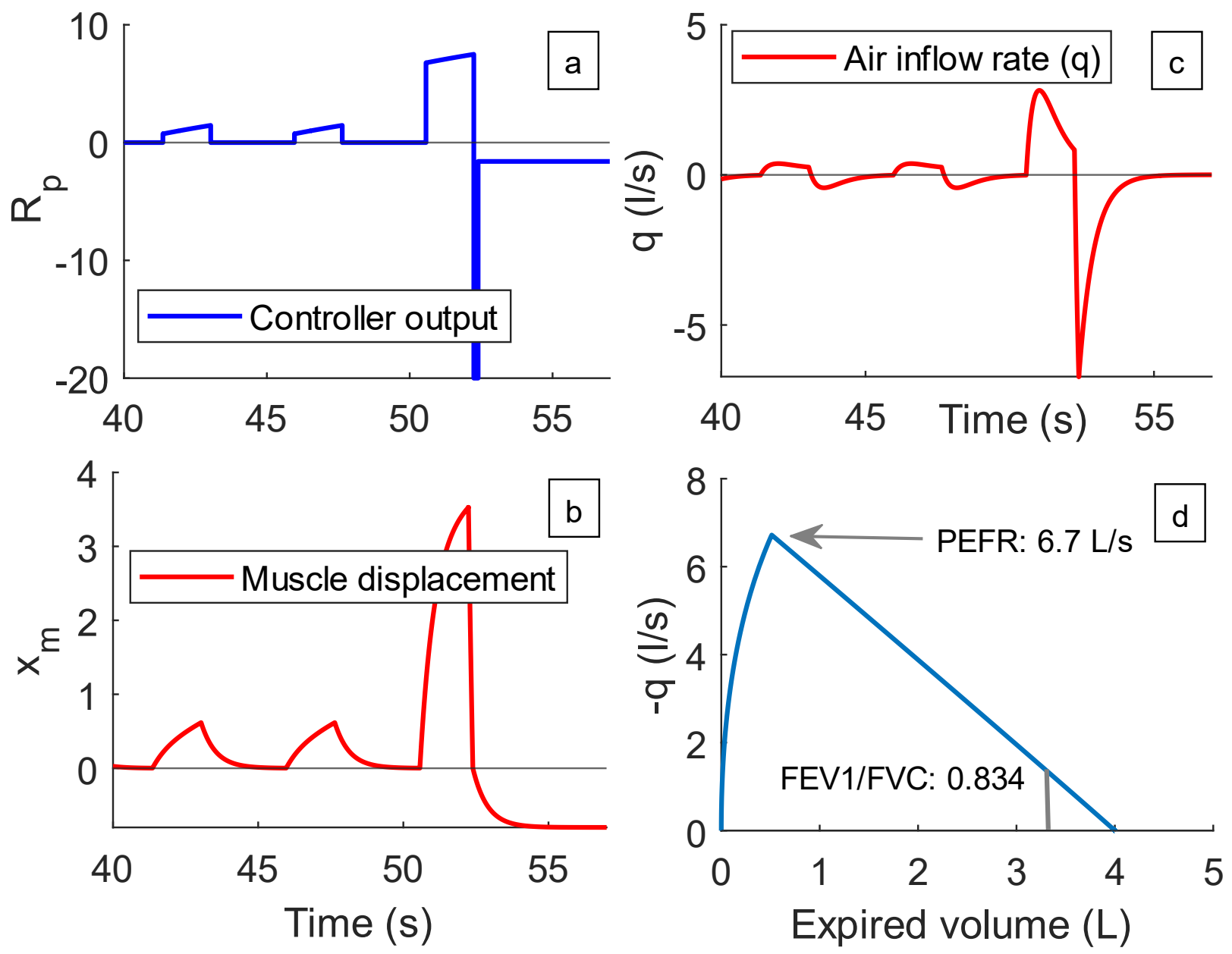

*Figure 14 (a) Changes in $R_p$ leading to changes in (b) muscle displacement and (c) air inflow rate. The spirogram is given in (d).*

## 3. Multivariable analysis of adaptations in COPD

This section describes the design of simulation case studies for performing a multi variable analysis of the different model adaptation in COPD: The steps involved are as follows:

1. We have three factors i.e. (i) airway resistance, (ii) chest elastance, and (iii) the resistance of the pulmonary arteries - that we would want to investigate the effects of, based on the response of the model.

2. Each factor has two levels, one corresponding to a healthy person and another representing a COPD patient.

3. Accounting for all combinations of the different factors, we have eight sets of parameters.

4. For each set of parameters, a deep breath was simulated to identify the spirometric variables.

5. In the next step, for each set of parameters, the model was simulated for (i) 1000 seconds under the condition of breathing normal air concentration and resting metabolic rate and (ii) 1000 seconds under an inspired oxygen concentration of 12%.

6. The values of the eight response variables that are good indicators of the COPD symptoms were recorded for each case. They are forced vital capacity (FVC), total lung capacity (TLC), PEFR (Peak Expiratory Flow Rate), $FEV1$ ratio, heart rate, pulmonary artery pressure ($P_{pa}$), respiratory rate, and minute ventilation.

7. Standardized t-tests were performed to identify whether the null hypothesis that the effect of each factor on the variable is zero can be rejected.

## 4. Parameters and variables

This section lists all the parameters and variables that appear in this paper. They are listed in separate tables for each engineering module explained in the model development section. As the parameter values are taken from multiple sources, the model response is indicative of the population average.

**Lung compartment and alveolar air**

| Symbol | Definition | Type | Value [Units] | Ref |
|---|---|---|---|---|
| $P_A$ | Alveolar pressure | Variable | [mmHg] | Ben-Tal (2008) |
| $V_A$ | Alveolar volume | Variable | [l] | Ben-Tal (2008) |
| $V_i$ | Inspired volume of the air | Variable | [l] | Ben-Tal (2008) |
| $V_D$ | Volume dead space | Parameter | [l] | Ben-Tal (2008) |
| $Q_A$ | Net flux of air to the alveoli | Variable | [l/s] | Ben-Tal (2008) |
| $f_{O2,al}$ | Molar fraction of O2 in the alveoli | Variable | [-] | Ben-Tal (2008) |
| $f_{O2,d}$ | Molar fraction of O2 in the volume dead space | Variable | [-] | Ben-Tal (2008) |
| $f_{CO2,al}$ | Molar fraction of CO2 in the alveoli | Variable | [-] | Ben-Tal (2008) |
| $f_{CO2,d}$ | Molar fraction of CO2 in the volume dead space | Variable | [-] | Ben-Tal (2008) |
| $f_{O2,i}$ | Molar fraction of O2 in the inspired air | Variable | [-] | Ben-Tal (2008) |
| $f_{CO2,i}$ | Molar fraction of CO2 in the inspired air | Variable | [-] | Ben-Tal |
| | | | | (2008) |
| $p_{O2,al}$ | Alveolar partial pressure of O2 | Variable | [mmHg] | Ben-Tal (2008) |
| $p_{CO2,al}$ | Alveolar partial pressure of CO2 | Variable | [mmHg] | Ben-Tal (2008) |
| $c_{O2,cap}$ | O2 Concentration in the lung capillaries | Variable | [$mg_{gas}/mg_{blood}$] | Ben-Tal (2008) |
| $p_{CO2,cap}$ | CO2 Partial pressure in the lung capillaries | Variable | [mmHg] | Ben-Tal (2008) |
| $c^{-}_{HCO3,cap}$ | Carbonic acid concentration in the lung capillaries | Variable | [mol/l] | Ben-Tal (2008) |
| $c_{O2,Pa}$ | O2 Concentration in the pulmonary arteries | Variable | [$mg_{gas}/mg_{blood}$] | - |
| $p_{CO2,Pa}$ | CO2 Partial pressure in the pulmonary artery | Variable | [mmHg] | - |
| $c^{-}_{HCO3,Pa}$ | Carbonic acid concentration in the pulmonary artery | Variable | [mol/l] | - |
| $q$ | Air flow through the airways | Variable | [l/s] | |
| $\sigma_{O2}$ | (Solubility of O2 in plasma) | Parameter | $1.4\times 10^{-6}$[mol/l.mmHg] | Ben-Tal (2006) |
| $\sigma_{CO2}$ | (Solubility of CO2 in plasma) | Parameter | $3.3\times 10^{-5}$[mol/l.mmHg] | Ben-Tal (2006) |
| $r_2$ | (Hydration reaction rate) | Parameter | 0.12 [1/s] | Ben-Tal (2006) |
| $l_2$ | (Dehydration reaction rate) | Parameter | $164\times 10^{3}$ [l/mol . s] | Ben-Tal (2006) |
| $T_h$ | (Threshold hemoglobin concentration) | Parameter | $2\times 10^{3}$ [mol/l] | Ben-Tal (2006) |
| $\delta$ | (Acceleration rate) | Parameter | $10^{1.9}$ [-] | Ben-Tal (2006) |
| $h$ | (concentration of $H^{+}$) | Parameter | $10^{-7.4}$ [mol/l] | Ben-Tal (2006) |
| $R$ | (Airways resistance to flow) | Parameter | 1 [mmHg. s /l] | Ben-Tal |
| | | | | (2006) |
| $p_w$ | (Vapor pressure of water) | Parameter | 47 [mmHg] | Ben-Tal (2006) |
| $D_{O2}$ | (Diffusion capacity O2) | Parameter | $3.5 \times 10^{-4}$ [l/s . mmHg] | Ben-Tal (2006) |
| $D_{CO2}$ | (Diffusion capacity CO2) | Parameter | $7.08 \times 10^{-3}$ [l/s . mmHg] | Ben-Tal (2006) |
| $V_{cap}$ | (Capillaries volume) | Parameter | 0.07 [l] | Ben-Tal (2006) |
| $E_T$ | (Chest elastance) | Parameter | 2.5 [mmHg/l] | Ben-Tal (2008) |
| $V_0$ | Volume of the lungs when it is unloaded | Parameter | 0 [l] | Ben-Tal (2008) |
| $K_{O2}$ | $O_2$ dissociation coefficient | Parameter | 0.2 [-] | Ellwein (2013) |
| $k_{o2}$ | $O_2$ dissociation coefficient | Parameter | 0.046 [-] | Ellwein (2013) |
| $K_{co2}$ | $CO_2$ dissociation coefficient | Parameter | 0.0065 [-] | Ellwein (2013) |
| $k_{co2}$ | $CO_2$ dissociation coefficient | Parameter | 0.244 [-] | Ellwein (2013) |
| $K_T$ | Equilibrium constant in the saturation function of hemoglobin | Parameter | $1.0 \times 10^4$ [l/mol] | Ben-Tal (2006) |
| $K_R$ | Equilibrium constant in the saturation function of hemoglobin | Parameter | $3.6 \times 10^6$ [l/mol] | Ben-Tal (2006) |
| $L$ | Equilibrium constant in the saturation function of hemoglobin | Parameter | $171.2 \times 10^6$ [-] | Ben-Tal (2006) |

**Respiratory sensors**

| Symbol | Definition | Type | Value [Units] | Ref |
|---|---|---|---|---|
| $p_{o2,s}$ | Blood partial pressure of O2 in the systemic arteries measured by arterial chemoreceptors | Variable | - | Christie (2014) |
| $p_{co2,s}$ | Blood partial pressure of CO2 in the Arteries Measured by arterial chemoreceptors | Variable | - | Christie (2014) |
| $c_{HCO3,s}^{-}$ | Concentration of Carbonic Acid (Representation of PH) measured by arterial chemoreceptors | Variable | - | Christie (2014) |
| $K_{O_2,s}$ | Sensor gain for oxygen | Parameter | 1 (healthy person) | Christie (2014) |
| $K_{CO_2,s}$ | Sensor gain for carbon dioxide | Parameter | 1 (healthy person) | Christie (2014) |
| $K_{HCO3^-,s}$ | Sensor gain for carbonic acid | Parameter | 1 (healthy person) | Christie (2014) |

**Respiratory control center**

| Symbol | Definition | Type | Value [Units] | Ref |
|---|---|---|---|---|
| $A$ | Average measure of spike rate in the neurons population | Variable | | Ben-Tal (2008) |
| $h_p$ | Inactivation gating of persistent sodium current | Variable | | Ben-Tal (2008) |
| $m_p$ | Activation gating of persistent sodium current | Variable | | Ben-Tal (2008) |
| $R_p(t)$ | Phrenic activity signal causing the lung muscle force | Variable | | Ben-Tal (2008) |
| $\tilde{g}_t$ | Controller parameter responsible for respiratory frequency | Variable | | Tuned using data in Guyton & Hall (2006) |
| $K_{ctrl}$ | Controller parameter responsible for respiratory amplitude | Variable | | Tuned using data in Guyton & Hall (2006) |
| $\tilde{g}_{nap}$ | (Activating rate constant of A) | Parameter | 133.33 [1/s] | Butera |
| | | | | (1999) |
| $\tilde{g}_L$ | (Inactivating rate constant of A) | Parameter | 98 [1/s] | Butera (1999) |
| $g_n$ | (nominal values of the control parameter $g_t$) | Parameter | 5-22 [1/s] | Butera (1999) |
| $\tilde{\theta}_{mp}$ | (Parameter affecting $\overline{m}_p$) | Parameter | 0.367 [-] | Butera (1999) |
| $\tilde{\sigma}_{mp}$ | (Parameter affecting $\overline{m}_p$) | Parameter | -0.033 [-] | Butera (1999) |
| $\tilde{\theta}_{hp}$ | (parameter affecting $h_p$) | Parameter | 0.313 [-] | Butera (1999) |
| $\tilde{\sigma}_{hp}$ | (parameter affecting $h_p$) | Parameter | 0.04 [-] | Butera (1999) |
| $\bar{\tau}_{hp}$ | (parameter affecting $h_p$) | Parameter | 10 [s] | Butera (1999) |
| $\tilde{E}_L$ | (parameter affecting the external drive of A) | Parameter | 0.212 [-] | Butera (1999) |
| $Tr_1$ | Threshold value in the ramp generation function | Parameter | 0.35 [-] | Ben-Tal (2008) |
| $Tr_2$ | Threshold value in the ramp generation function | Parameter | 0.35 [-] | Ben-Tal (2008) |
| $Tr_3$ | Threshold value in the ramp generation function | Parameter | 0.3 [-] | Ben-Tal (2008) |
| $Tr_4$ | Threshold value in the ramp generation function | Parameter | 0.001 [-] | Ben-Tal (2008) |

**Lung Muscles (Actuator)**

| Symbol | Definition | Type | Value [Units] | Ref |
|---|---|---|---|---|
| $P_L$ | Pleural pressure | Variable | [mmHg] | Ben-Tal (2008) |
| $x_m$ | Muscle displacement | Variable | [m] | Ben-Tal (2008) |
| $P_m$ | Mouth pressure | Parameter | 760 [mmHg] | Ben-Tal (2008) |
| $P_{L0}$ | Difference between atmospheric and pleural pressure | Parameter | 4.5 [mmHg] | Ben-Tal (2008) |
| $k_1$ | Recoil rate constant of muscle | Parameter | 2 [1/s] | Ben-Tal (2008) |
| $k_2$ | Conversion constant | Parameter | 1 [m/s] | Ben-Tal (2008) |
| $k_p$ | Conversion constant | Parameter | 2.5 [mmHg/m] | Ben-Tal (2008) |

**Cardiovascular system**

| Symbol | Definition | Type | Value [Units] | Ref |
|---|---|---|---|---|
| $q_{pv}$ | Pulmonary vein blood flow rate | Variable | [ml/s] | Ellwein (2013) |
| $P_{pv}$ | Pulmonary vein pressure | Variable | [mmHg] | Ellwein (2013) |
| $C_{pv}$ | Pulmonary venous Compliance | Parameter | 22.4 [ml/mmHg] | Ellwein (2013) |
| $R_p$ | Pulmonary Resistance | Parameter | 0.198 [mmHg.s/ml] | Ellwein (2013) |
| $R_{mv}$ | Mitral valve Resistance | Variable | [mmHg.s/ml] | Ellwein (2013) |
| $R_{\mathrm{mv},o}$ | Open mitral valve resistance | Parameter | 0.001 [mmHg.s/ml] | Ellwein (2013) |
| $R_{\mathrm{mv},c}$ | Closed mitral valve resistance | Parameter | 20 [mmHg.s/ml] | Ellwein (2013) |
| $q_{lv}$ | Left ventricle blood flow rate | Variable | [ml/s] | Ellwein (2013) |
| $V_{lv}$ | Left ventricular volume | Variable | [ml] | Ellwein (2013) |
| $E_{lv}$ | Left ventricular elastance | Variable | [mmHg/ml] | Ellwein (2013) |
| $E_{D,l}$ | Left ventricular diastolic elastance | Parameter | 0.0135 [mmHg/ml] | Ellwein (2013) |
| $E_{s,l}$ | Left ventricular systolic elastance | Parameter | 0.849 [mmHg/ml] | Ellwein (2013) |
| $q_{sa}$ | Systemic artery blood flow rate | Variable | [ml/s] | Ellwein (2013) |
| $P_{Sa}$ | Systemic arteries pressure | Variable | [mmHg] | Ellwein (2013) |
| $C_{sa}$ | Systemic arteries Compliance | Parameter | 2.4 [ml/mmHg] | Ellwein (2013) |
| $R_s$ | Systemic Resistance | Parameter | 1.08 [mmHg.s/ml] | Ellwein (2013) |
| $R_{av}$ | Aortic Valve Resistance | Variable | [mmHg.s/ml] | Ellwein (2013) |
| $R_{\text{av},o}$ | Open aortic valve valve resistance | Parameter | 0.001 [mmHg.s/ml] | Ellwein (2013) |
| $R_{\text{av},c}$ | Closed aortic valve valve resistance | Parameter | 20 [mmHg.s/ml] | Ellwein (2013) |
| $c_{CO_2,Sv}$ | CO2 concentration in the systemic vein | Variable | [mg/ml] | Ellwein (2013) |
| $c_{CO_2,\text{Sa}}$ | CO2 concentration in the systemic arteries | Variable | [mg/ml] | Ellwein (2013) |
| $M_{S,CO2}$ | Body CO2 tissue metabolism | Variable | [mg/s] | Batzel (2005) |
| $M_{S,O2}$ | Body O2 tissue metabolism | Variable | [mg/s] | Batzel (2005) |
| $q_{sv}$ | Systemic vein blood flow rate | Variable | [ml/s] | Batzel (2005) |
| $P_{Sv}$ | Systemic vein pressure | Variable | [mmHg] | Batzel (2005) |
| $C_{sv}$ | Systemic venous Compliance | Parameter | 5.95 [ml/mmHg] | Batzel (2005) |
| $R_s$ | Systemic Resistance | Parameter | 1.08 [mmHg . s/ml] | Batzel (2005) |
| $R_{\mathrm{tv}}$ | Tricuspid valve resistance | Variable | [mmHg . s/ml] | Batzel (2005) |
| $R_{\mathrm{tv},o}$ | Open Tricuspid valve resistance | Parameter | 0.001 [mmHg . s/ml] | Batzel (2005) |
| $R_{\mathrm{tv},c}$ | Closed Tricuspid valve resistance | Parameter | 20 [mmHg . s/ml] | Batzel (2005) |
| $q_{rv}$ | Right ventricle blood flow rate | Variable | [ml/s] | Batzel (2005) |
| $V_{\mathrm{rv}}$ | Right ventricular volume | Variable | [ml] | Batzel (2005) |
| $E_{\mathrm{rv}}$ | Right ventricular elastance | Variable | [mmHg/ml] | Batzel (2005) |
| $E_{D,r}$ | Right ventricular diastolic elastance | Parameter | 0.0667 [mmHg/ml] | Batzel (2005) |
| $E_{s,r}$ | Right ventricular systolic elastance | Parameter | 3.38 [mmHg/ml] | Batzel (2005) |
| $q_{pa}$ | Pulmonary artery blood flow rate | Variable | [ml/s] | Batzel (2005) |
| $P_{pa}$ | Pulmonary artery pressure | Variable | [mmHg] | Batzel (2005) |
| $C_{pa}$ | Pulmonary arterial Compliance | Parameter | 4.52 [ml/mmHg] | Batzel (2005) |
| $R_p$ | Pulmonary Resistance | Parameter | 0.198 [mmHg.s/ml] | Batzel (2005) |
| $R_{pv}$ | Pulmonary valve Resistance | Variable | [mmHg.s/ml] | Batzel (2005) |
| $R_{\mathrm{pv},o}$ | Open $pv$ resistance | Parameter | 0.001 [mmHg.s/ml] | Batzel (2005) |
| $R_{\mathrm{pv},c}$ | Closed $pv$ resistance | Parameter | 20 [mmHg.s/ml] | Batzel (2005) |
| $H$ | Mean Heart Rate | Variable | [beat/min] | Batzel (2005) |
| $T_{M,f}$ | Fraction of time to end-systolic to the heart time period | Parameter | 0.35 | Heldt (2004) |
| $T_R, f$ | Fraction of ventricular relaxation time to the heart time period | Parameter | 0.35 | Heldt (2004) |
| $T$ | Heart Time Period | Variable | [min] | Batzel (2005) |

**Polynomial coefficients of Equation S.32**

| Symbol | Value |
|---|---|
| $W$ | 0.014 |
| $h_1$ | 3.23E-6 |
| $i_1$ | -4.46E-4 |
| $j_1$ | 2.25E-2 |
| $p_1$ | -4.79E-1 |
| $q_1$ | 4.37 |
| $r$ | -43 |
| $f$ | -0.003 |
| $g$ | 98 |
| $s$ | 2.3 |

## 5. Initial conditions

Initial condition for the cardio-respiratory model

| Symbol | Meaning | Initial value |
|---|---|---|
| $P_A$ | Alveolar pressure | 760 [mmHg] |
| $f_{O2,al}$ | The molar fraction of O2 in the alveoli | 0.2 [-] |
| $f_{CO2,al}$ | The molar fraction of CO2 in the alveoli | 0.0095 [-] |
| $p_{O2,Sa}$ | O2 partial pressure in the systemic arteries | 125 [mmHg] |
| $p_{CO2,Sa}$ | CO2 partial pressure in the systemic | 20 [mmHg] |
| | arteries | |
| $c^-_{HCO3,Sa}$ | bicarbonate concentration in the systemic arteries | 0.028 [mol/l] |
| $A$ | Average measure of spike rate in the neurons population | 0.34 [-] |
| $h_p$ | Inactivation gating of persistent sodium current | 0.33 [-] |
| $x_m$ | Muscle displacement | 0 [-] |
| $P_{pv}$ | Pulmonary vein pressure | 3.3 [mmHg] |
| $P_{Sv}$ | Systemic vein pressure | 6.6 [mmHg] |
| $P_{Sa}$ | Systemic arteries pressure | 79.5 [mmHg] |
| $P_{pa}$ | Pulmonary artery pressure | 20 [mmHg] |
| $V_{lv}$ | Left ventricular volume | 312 [ml] |
| $V_{rv}$ | Right ventricular volume | 100 [ml] |

## 6. DATA USED FOR DEFINING CONTROLLER PARAMETERS $\tilde{g}_t$ AND $K_{ctrl}$

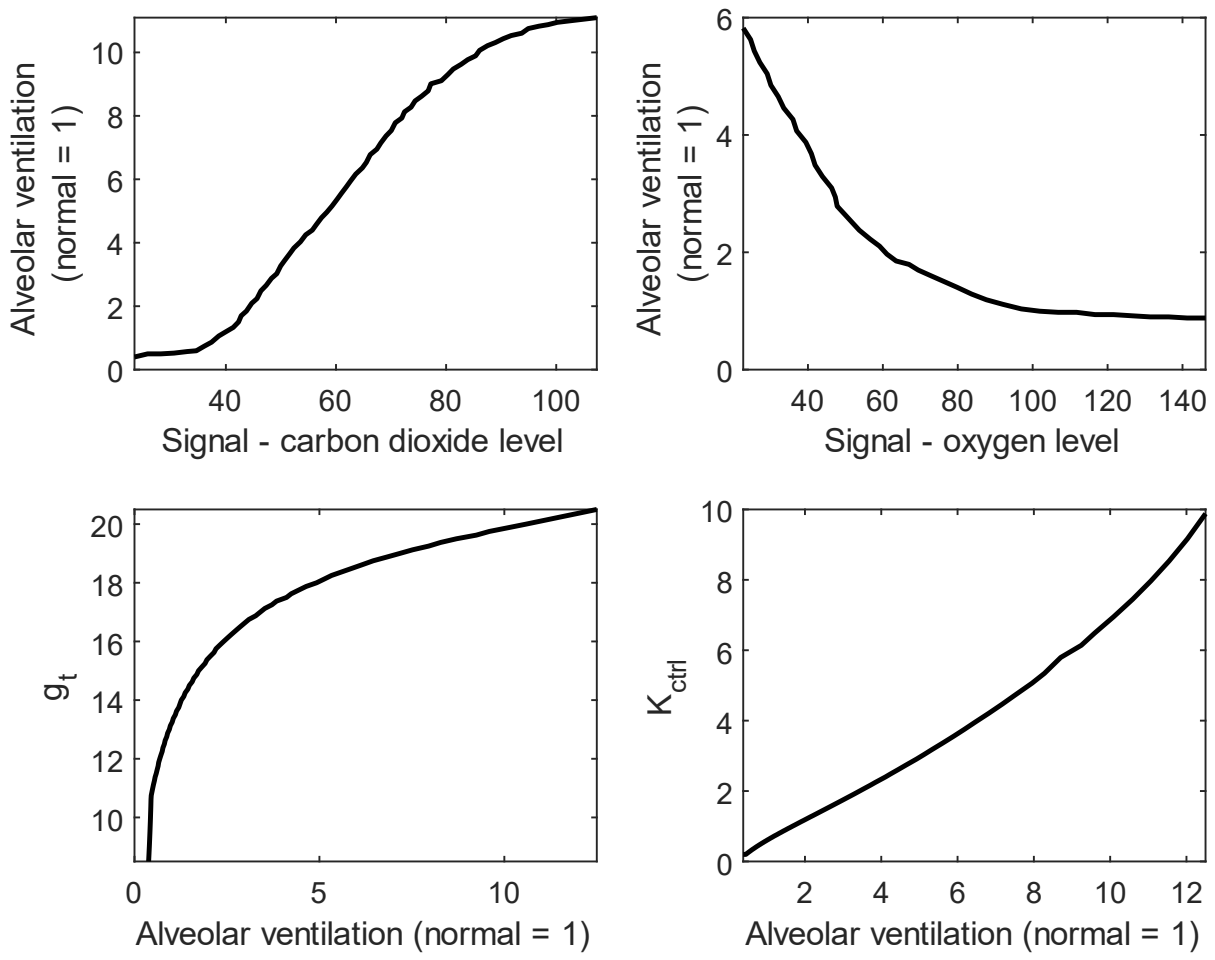

*Figure 15 The data used for connecting the blood gas measurements to the controller parameters, $\tilde{g}_t$ and $K_{ctrl}$ [39]*